\documentclass[11pt,a4paper]{article} 
\pdfoutput=1
\usepackage{jheppub,hyperref,mdwlist}
\usepackage{amsmath}
\usepackage{graphicx}
\usepackage{subfigure}
\usepackage{amssymb}
\usepackage{xcolor}
\usepackage{multirow}
\usepackage{cancel}
\usepackage{color}
\usepackage{listings}
\usepackage{xcolor}
\usepackage{float}
\usepackage{slashed}

\usepackage{amsmath}%\ge=\geqslant; \le=\leqslant
\usepackage{wrapfig}
\usepackage{cleveref}

\newcommand{\be}{\begin{equation}}
\newcommand{\ee}{\end{equation}}
\newcommand{\beq}{\begin{equation}}
\newcommand{\eeq}{\end{equation}}
\newcommand{\bea}{\begin{eqnarray}}
\newcommand{\eea}{\end{eqnarray}}
\newcommand{\besp}{\begin{equation}\begin{split}}
\newcommand{\eesp}{\end{split}\end{equation}}

\newcommand{\nn}{\nonumber}

\newcommand{\Eq}[1]{Eq.~(\ref{#1})}
\newcommand{\Dfbd}{\mathord{\buildrel{\lower3pt\hbox{$\scriptscriptstyle\leftrightarrow$}}\over {D}_{\mu}}}
\newcommand{\ave}[1]{\left\langle #1\right\rangle}

\def\mG{\mathcal{G}}

\def\mL{\mathcal{L}}
\def\mM{\mathcal{M}}

\def\mO{\mathcal{O}}

\def\Z{\mathbb{Z}}

\def\0{\textbf{0}}
\def\1{\textbf{1}}
\def\2{\textbf{2}}
\def\3{\textbf{3}}
\def\4{\textbf{4}}
\def\5{\textbf{5}}
\def\6{\textbf{6}}
\def\7{\textbf{7}}
\def\8{\textbf{8}}
\def\9{\textbf{9}}

\def\hc{\text{h.c.}}

\begin{document}

\title{Probing superheavy dark matter with gravitational waves}

\author[a,b]{Ligong Bian,}
\author[c]{Xuewen Liu,}
\author[d,e,1]{Ke-Pan Xie}
\footnotetext[1]{Corresponding author.}

\affiliation[a]{Department of Physics, Chongqing University, Chongqing 401331, China}
\affiliation[b]{Chongqing Key Laboratory for Strongly Coupled Physics, Chongqing 401331, China}
\affiliation[c]{Department of Physics, Yantai University, Yantai 264005, China}
\affiliation[d]{Center for Theoretical Physics, Department of Physics and Astronomy, Seoul National University, Seoul 08826, Korea}
\affiliation[e]{Department of Physics and Astronomy, University of Nebraska, Lincoln, NE 68588, USA}

\emailAdd{lgbycl@cqu.edu.cn}
\emailAdd{xuewenliu@ytu.edu.cn}
\emailAdd{kepan.xie@unl.edu}

\abstract{

We study the superheavy dark matter (DM) scenario in an extended $B-L$ model, where one generation of right-handed neutrino $\nu_R$ is the DM candidate. If there is a new lighter sterile neutrino that co-annihilate with the DM candidate, then the annihilation rate is exponentially enhanced, allowing a DM mass much heavier than the Griest-Kamionkowski bound ($\sim10^5$ GeV). We demonstrate that a DM mass $M_{\nu_R}\gtrsim10^{13}$ GeV can be achieved. Although beyond the scale of any traditional DM searching strategy, this scenario is testable via gravitational waves (GWs) emitted by the cosmic strings from the $U(1)_{B-L}$ breaking. Quantitative calculations show that the DM mass $\mathcal{O}(10^9-10^{13}~{\rm GeV})$ can be probed by future GW detectors.

}

\maketitle
\flushbottom

\section{Introduction}

The freeze-out of weakly interacting massive particles (WIMPs)~\cite{Lee:1977ua} has been the most popular explanation for the particle origin of dark matter (DM) for decades. In this paradigm, the DM relic abundance is determined by~\cite{Kolb:1990vq,Bertone:2004pz,Lisanti:2016jxe}
\be\label{freeze-out}
\Omega_{\rm DM}h^2\sim0.1\times\frac{1~{\rm pb}}{\ave{\sigma_{\rm ann.}v}}\sim0.1\times\left(\frac{0.01}{\alpha_{\rm DM}}\right)^2\left(\frac{M_{\rm DM}}{100~{\rm GeV}}\right)^2,
\ee
with $\sigma_{\rm ann.}$ being the pair annihilation cross section of DM particles to the Standard Model (SM) particles, $v$ the relative velocity, and $\ave{...}$ the thermal average. In the second approximate equality we have used $\ave{\sigma_{\rm ann.}v}\sim\alpha_{\rm DM}^2/M_{\rm DM}^2$ where $\alpha_{\rm DM}$ is the finite structure constant of the coupling between the dark and SM sectors. \Eq{freeze-out} shows that if the dark matter is at electroweak (EW) scale and its coupling is of the order of the EW coupling, then the freeze-out mechanism can explain the observed DM density $\Omega_{\rm DM}h^2\approx0.12$~\cite{Planck:2018vyg,ParticleDataGroup:2020ssz}. This is the so-called ``WIMP miracle'', which has motivated enormous efforts to search for EW scale WIMPs via direct~\cite{Schumann:2019eaa}, indirect~\cite{Gaskins:2016cha} and collider~\cite{Kitano:2010fa} experiments.

However, WIMP mass deviating from EW scale is possible. For example if $\alpha_{\rm DM}\gg0.01$, then $M_{\rm DM}\gg100$ GeV can yield the correct DM abundance. Since $\alpha_{\rm DM}$ has an upper limit $\sim4\pi$ set by the unitarity bound, $M_{\rm DM}$ also has an upper limit, which is $\sim300$ TeV derived from the partial wave analysis, known as the Griest-Kamionkowski (GK) bound~\cite{Griest:1989wd}.\footnote{This bound applies to the elementary particle DM. If the DM is a composite object, another bound $R_{\rm DM}\gtrsim7.5\times10^{-7}$ fm applies to the DM size~\cite{Griest:1989wd}.} It is known that DM can be heavier than the GK bound in case of non-thermal dynamics, non-standard cosmological history~\cite{Kolb:1998ki,Hui:1998dc,Chung:1998rq,Chung:2001cb,Harigaya:2014waa,Davoudiasl:2015vba,Randall:2015xza,Harigaya:2016vda,Berlin:2016vnh,Berlin:2016gtr,Bramante:2017obj,Hamdan:2017psw,Cirelli:2018iax,Babichev:2018mtd,Hashiba:2018tbu,Hooper:2019gtx,Davoudiasl:2019xeb,Chanda:2019xyl} or the first-order cosmic phase transition~\cite{Baker:2019ndr,Chway:2019kft,Marfatia:2020bcs,Baldes:2020kam,Azatov:2021ifm}. However, as proposed in Refs.~\cite{Berlin:2017ife,Kramer:2020sbb}, DM mass beyond the GK bound is also possible within the thermal freeze-out framework, as long as there is a lighter unstable species that co-annihilates with the DM candidate (dubbed as the ``zombie collision''~\cite{Kramer:2020sbb}) and exponentially enhances the interaction rate. In other words, for the same coupling, such an annihilation allows an exponentially heavier DM mass compared to the normal DM pair annihilation scenario. The extra entropy produced from the late time decay of the lighter species further dilutes the DM density, leading to an even higher DM mass upper limit that can reach $10^{16}$ GeV~\cite{Berlin:2017ife}.

While the above zombie annihilation mechanism is theoretically appealing, it is very challenging to probe such superheavy DM via the traditional direct, indirect or collider experiments. In this article, we propose a zombie annihilation mechanism that is  associated with the breaking of a $U(1)$ symmetry, which leads to the formation of cosmic strings that can be detected via the gravitational wave (GW) signals at current or future GW detectors. As a benchmark, we study an extended $B-L$ model, where one generation of the right-handed neutrino (RHN) is the DM candidate, and a new Dirac sterile neutrino serves as the lighter species for co-annihilation. Section~\ref{sec:model} introduces the model, while Section~\ref{sec:freeze-out} is devoted to the calculation of thermal freeze-out and DM relic abundance. We then investigate the corresponding cosmic strings and GW signals in Section~\ref{sec:gws}, where the correlation to the DM scenario is obtained, and the recent NANOGrav excess is also commented. The conclusion will be given in Section~\ref{sec:conclusion}.

\section{The extended $B-L$ model}\label{sec:model}

We begin with the $B-L$ model~\cite{Davidson:1978pm,Marshak:1979fm,Mohapatra:1980qe,Davidson:1987mh}, which gauges the $U(1)_{B-L}$ group, with $B$ and $L$ being the baryon and lepton number, respectively. In the $B-L$ model, three generations of Majorana RHNs $\nu_R^i$ (with $B-L=-1$) are introduced for gauge anomaly cancellation. The new gauge boson of $U(1)_{B-L}$ is denoted as $Z'$, and a complex scalar field $\Phi$ (with $B-L=2$) is introduced to break the $U(1)_{B-L}$ and provide mass for $Z'$. The relevant Lagrangian reads
\be\label{LB-L}\begin{split}
\mL_{B-L}=&~\sum_{i}\bar\nu_R^ii\slashed{D}\nu_R^i-\frac12\sum_{i,j}\left(\lambda_R^{ij}\bar\nu_R^{i,c}\Phi\nu_R^j+\hc\right)
-\sum_{i,j}\left(\lambda_D^{ij}\bar\ell_L^i\tilde H\nu_R^j+\hc\right)\\
&~+D_\mu\Phi^\dagger D^\mu\Phi-\lambda_\phi\left(|\Phi|^2-\frac{v_\phi^2}{2}\right)^2-\frac14Z'_{\mu\nu}Z'^{\mu\nu},
\end{split}\ee
where $i$, $j$ are generation indices and $D_\mu=\partial_\mu-ig_{B-L}XZ_\mu'$ is the gauge covariant derivative with $X$ being the corresponding $B-L$ number. The scalar potential triggers a spontaneous symmetry breaking at $\ave{|\Phi|}=v_\phi/\sqrt{2}$. If we parametrize $\Phi$ as $(\phi+i\eta)/\sqrt{2}$, then after the $U(1)_{B-L}$ breaking $\eta$ is absorbed as the longitudinal mode of $Z'$, and the particles obtain the following masses
\be
M_{Z'}=2g_{B-L}v_\phi,\quad M_{\nu_R}^{ij}=\lambda_R^{ij}\frac{v_\phi}{\sqrt2},\quad M_\phi=\sqrt{2\lambda_\phi}v_\phi.
\ee
The Lagrangian (\ref{LB-L}) can elegantly explain the extremely small mass of the SM neutrinos and the matter-antimatter asymmetry of the Universe via Type-I seesaw~\cite{Minkowski:1977sc} and leptogenesis~\cite{Fukugita:1986hr}, respectively. Especially, the explanation of neutrino mass $\sim0.1$ eV~\cite{Planck:2018vyg,KATRIN:2019yun} prefers superheavy RHNs, since the seesaw mechanism requires $M_{\nu_R}\sim \lambda_D\lambda_D^\dagger\times10^{14}~{\rm GeV}$.

\begin{table}\small\renewcommand\arraystretch{1.5}\centering
\begin{tabular}{c|c|c|c|c|c}\hline\hline
 & $\nu_R^{1,2}$ & $\nu_R^3$ & $\Phi$ & $\psi$ & $S$ \\ \hline
$B-L$ & $-1$ & $-1$ & 2 & $-1$ & 0 \\ \hline
 $\Z_2$ & 1 & $-1$ & 1 & $-1$ & 1 \\ \hline\hline
\end{tabular} 
\caption{Quantum numbers of the BSM particles. All particles listed here are singlets under the SM gauge group.}\label{tab:quantum_number}
\end{table}

For the sake of a DM candidate, we adjust \Eq{LB-L} by assigning a $\Z_2$ symmetry, under which the third generation RHN is odd while all other particles are even. Consequently, $\lambda_D^{i3}=0$ for $i=1$, 2, 3 and $\lambda_R^{j3}=0$ for $j=1$, 2, making $\nu_R^3$ free from the $\nu_R^3\to\ell H$ decay and hence can be a DM candidate. This setup is generally adopted in the $\nu_R$-DM scenarios~\cite{Okada:2010wd,Okada:2012sg,Okada:2016tci,Okada:2016gsh,Okada:2018ktp,Borah:2018smz}.\footnote{Two generations of $\Z_2$-even RHNs are already sufficient to explain the neutrino oscillation data and realize leptogenesis, see the recent review~\cite{Xing:2020ald} and the references therein.} To realize the zombie co-annihilation, we further extend the model by introducing a Dirac sterile neutrino $\psi$ (with $B-L=-1$) and a gauge singlet real scalar mediator $S$.\footnote{See Ref.~\cite{Bian:2019szo} for the low-scale phase transition study in the complex singlet extended $U(1)_{B-L}$ model.} The quantum numbers of the particles beyond the SM (BSM) are summarized in Table~\ref{tab:quantum_number}. Since hereafter we only discuss the DM candidate $\nu_R^3$, for simplicity we will denote $\nu_R^3$ as $\nu_R$ and $\lambda_R^{33}$ as $\lambda_R$. The Lagrangian relevant for DM reads
\be\label{LDM}
\mL_{\rm DM}=\bar\psi\left(i\slashed{D}-M_\psi\right)\psi+\frac12\partial_\mu S\partial^\mu S-\frac12M_S^2S^2+\left(\lambda_1 S\bar\psi\nu_R+\hc\right)+\lambda_2 S\bar\psi\psi,
\ee
which provides the zombie collisions
\be\label{zombie}
\nu_R+\psi\to \psi+\psi,\quad \nu_R+\psi\to\psi+\bar\psi,
\ee
and their charge conjugations via the exchange of an $S$ mediator. If we name $\nu_R$ as a ``survivor'' and $\psi$ as a ``zombie'', then \Eq{zombie} is infecting a survivor to a zombie, and that is why it is called a zombie collison~\cite{Kramer:2020sbb}. As we will see in the next section, for $M_\psi<M_{\nu_R}=\lambda_Rv_\phi/\sqrt{2}$, the annihilation rate is exponentially enhanced and $M_{\nu_R}$ can reach $10^{13}$ GeV while still generating the correct DM relic abundance.

We work in the parameter space
\be\label{mass_relation}
M_{\nu_R}+M_\psi<M_S,\quad M_\psi<M_{\nu_R}<3M_\psi,
\ee
so that the DM decay channels $\nu_R\to \psi S/\bar\psi S$ or $\nu_R\to\psi\psi\bar\psi$ are kinematically forbidden, while the mediator $S$ can decay to pairs of $\bar\psi\nu_R,\psi\nu_R$, and $\psi\bar\psi$, with the width
\be
\Gamma_S=\Gamma_{S\to\bar\psi\nu_R}+\Gamma_{S\to\psi\nu_R}+\Gamma_{S\to\psi\bar\psi},
\ee
where
\be\begin{split}
\Gamma_{S\to\bar\psi\nu_R}=\Gamma_{S\to\bar\psi\nu_R}=&~\frac{\lambda_1^2}{16\pi}M_S\left(1-a_R-a_\psi\right)\sqrt{1-2\left(a_R+a_\psi\right)+\left(a_R-a_\psi\right)^2},\\
\Gamma_{S\to\psi\bar\psi}=&~\frac{\lambda_2^2}{8\pi}M_S\left(1-4a_\psi\right)^{3/2},
\end{split}\ee
and $a_R\equiv M_{\nu_R}^2/M_S^2$, $a_\psi\equiv M_\psi^2/M_S^2$.

The final essential ingredient of the zombie mechanism is an appropriate decay channel for the Dirac sterile neutrino $\psi$. After the freeze-out of $\nu_R$, $\psi$ needs to decay into the SM particles, otherwise itself is a WIMP DM candidate that satisfies the GK bound, and hence $\nu_R$ cannot exceed the GK bound due to the $M_{\nu_R}<3M_\psi$ constraint. Since $\psi$ is odd under the $\Z_2$ while the SM particles are even, such $\psi$ decay must involve some $\Z_2$-breaking interactions, which can also result in the decay of the DM candidate $\nu_R$. To make $\nu_R$ sufficiently long-lived (with a lifetime longer than $10^{27}$ s, as required by the diffuse gamma-ray spectrum~\cite{Cirelli:2012ut,Essig:2013goa,Blanco:2018esa}), the breaking of $\Z_2$ should be mediated by either extremely small couplings or extremely high scales with some level of fine-tuning.\footnote{The word ``fine-tuning'' is also in the sense that we assume only $\psi$ directly participates in the $\Z_2$-breaking interactions.} For the former case, we can have an interacting vertex like $y_\psi\bar\ell_LH\psi_R$ with $|y_\psi|\ll1$; while for the latter case, a dimension-6 operator can be induced by exchanging a new heavy color triplet scalar
\be\label{L6}
\mL_6=\frac{1}{\Lambda^2}\sum_{i,j,j'}(\bar\psi^c u_R^i)(\bar d_R^{j,c}d_R^{j'})+\hc,
\ee
where $i$, $j$ and $j'$ are generation indices and $j\neq j'$. The color indices of quarks are implicitly summed up in a completely asymmetric way via the Levi-Civita symbol to yield a color singlet. We will take \Eq{L6} as an example for the further study, but keep in mind that the zombie DM mechanism holds as long as $\psi$ can decay via a tiny $\Z_2$-breaking interaction. According to \Eq{L6}, the decay width of $\psi$ is
\be
\Gamma_\psi=\Gamma_{\psi\to 3j}=\frac{1}{1024\pi^3}\frac{M_\psi^5}{\Lambda^4},
\ee
assuming an exclusive $uds$-quark final state. Note that \Eq{L6} also allows the decay $\nu_R\to\psi\psi^{(*)}\bar\psi^{(*)}\to 9j$ via one or two off-shell $\psi$'s, and hence a high $\Lambda$ is required to keep $\nu_R$ sufficiently long-lived. On the other hand, the same high $\Lambda$ also results in the late time decay of $\psi$, which can produce entropy to dilute the $\nu_R$ abundance.

\section{The superheavy $\nu_R$ dark matter}\label{sec:freeze-out}

\subsection{Thermal freeze-out}

Assume the reheating temperature after inflation is higher than $M_{\nu_R}$ and hence both $\nu_R$ and $\psi$ are originally in thermal equilibrium with each other and the SM particles, and the number densities are described by
\be
n_\alpha^{\rm eq}=2\int\frac{d^3p}{(2\pi)^3}e^{-E_\alpha/T}=2\times\frac{M_\alpha^2T}{2\pi^2}K_2\left(\frac{M_\alpha}{T}\right),
\ee
where $\alpha=\nu_R$, $\psi$ or $\bar\psi$, $K_2$ is the modified Bessel function of the second kind, and the factor 2 is for spin degeneracy. 
As the Universe expands and cools down, the reaction rate between $\nu_R$ and $\psi$ (see \Eq{zombie}) drops. When the interaction rate is lower than the Universe expansion rate, $\nu_R$ deviates from the chemical equilibrium and eventually freeze-out to be the DM candidate. The freeze-out process can be characterized by a set of Boltzmann equations, which are discussed in detail below.

Consider the radiation dominated era that the energy and entropy densities are respectively
\be
\rho=\frac{\pi^2}{30}g_*T^4,\quad s=\frac{2\pi^2}{45}g_*T^3,
\ee
with $g_*$ being the number of relativistic degrees of freedom which we take to be the SM value 106.75. The Hubble constant can be solved as
\be
H=\left(\frac{8\pi}{3M_{\rm Pl}^2}\frac{\pi^2}{30}g_*T^4\right)^{1/2}=2\pi\sqrt{\frac{\pi g_*}{45}}\frac{T^2}{M_{\rm Pl}},
\ee
with $M_{\rm Pl}=1.22\times10^{19}$ GeV the Planck scale. Define the dimensionless parameter $z=M_{\nu_R}/T$, then $s=s_{\nu_R}/z^3$ and $H=H_{\nu_R}/z^2$, where
\be
s_{\nu_R}\equiv s|_{T=M_{\nu_R}},\quad H_{\nu_R}\equiv H|_{T=M_{\nu_R}}.
\ee
Finally, we define the particle abundance as $Y_\alpha=n_\alpha/s$, i.e. the ratio of number density to the entropy density, and hence the abundances of the equilibrium distributions are
\be
Y_{\nu_R}^{\rm eq}=\frac{45z^2}{2\pi^4g_*}K_2(z),\quad
Y_\psi^{\rm eq}=\frac{45z^2}{2\pi^4g_*}\left(\frac{M_\psi}{M_{\nu_R}}\right)^2K_2\left(\frac{M_\psi}{M_{\nu_R}}z\right).
\ee

Under above conventions, the Boltzmann equations of $\nu_R$ and $\psi$ can be expressed as a set of ordinary differential equations,
\begin{small}\bea\label{Boltzmann_1}
\frac{s_{\nu_R} H_{\nu_R}}{z^4}\frac{dY_{\nu_R}}{dz}&=&
-\frac{Y_\psi}{Y_\psi^{\rm eq}}\left(\frac{Y_{\nu_R}}{Y_{\nu_R}^{\rm eq}}-\frac{Y_\psi}{Y_\psi^{\rm eq}}\right)\left(2\gamma_{\nu_R\psi\to\psi\psi}+2\gamma_{\nu_R\bar\psi\to\psi\bar\psi}\right)\\
&&-\left(\frac{Y_{\nu_R}^2}{(Y_{\nu_R}^{\rm eq})^2}-\frac{Y_\psi^2}{(Y_\psi^{\rm eq})^2}\right)2\left(\gamma_{\nu_R\nu_R\to\psi\bar\psi}+2\gamma_{\nu_R\nu_R\to\psi\psi}\right)
-\left(\frac{Y_{\nu_R}^2}{(Y_{\nu_R}^{\rm eq})^2}-1\right)2\gamma_{\nu_R\nu_R\to f\bar f},\nn\\
\label{Boltzmann_2}
\frac{s_{\nu_R}H_{\nu_R}}{z^4}\frac{dY_\psi}{dz}&=&\frac{Y_\psi}{Y_\psi^{\rm eq}}\left(\frac{Y_{\nu_R}}{Y_{\nu_R}^{\rm eq}}-\frac{Y_\psi}{Y_\psi^{\rm eq}}\right)\left(\gamma_{\nu_R\psi\to\psi\psi}+\gamma_{\nu_R\bar\psi\to\psi\bar\psi}\right)\\
&&+\left(\frac{Y_{\nu_R}^2}{(Y_{\nu_R}^{\rm eq})^2}-\frac{Y_\psi^2}{(Y_\psi^{\rm eq})^2}\right)\left(\gamma_{\nu_R\nu_R\to\psi\bar\psi}+2\gamma_{\nu_R\nu_R\to\psi\psi}\right)
-\left(\frac{Y_\psi^2}{(Y_\psi^{\rm eq})^2}-1\right)\gamma_{\psi\bar\psi\to f\bar f},\nn
\eea\end{small}
and $Y_{\nu_R}^\infty=Y_{\nu_R}(z\to\infty)$ is the freeze-out relic abundance of the DM candidate $\nu_R$. Here the $\gamma_{ab\to cd}$'s are the corresponding interaction rates,
\be
\gamma_{ab\to cd}=n_a^{\rm eq}n_b^{\rm eq}\ave{\sigma_{ab\to cd}v},
\ee
and the detailed definitions can be found in Appendix~\ref{app:rates}. Since $M_{\nu_R}>M_\psi$, the abundance of $\psi$ is higher than $\nu_R$, i.e. $n_\psi^{\rm eq}\sim \left(n_{\nu_R}^{\rm eq}\right)^{M_{\nu_R}/M_\psi}$, leading to an exponentially enhanced annihilation rate compared with the WIMP scenario. The factor ``2'' in front of $\gamma_{\nu_R\psi\to\psi\psi}$, $\gamma_{\nu_R\bar\psi\to\psi\bar\psi}$ and $\gamma_{\nu_R\nu_R\to\psi\psi}$ comes from charge conjugation (e.g. $\gamma_{\nu_R\bar\psi\to\bar\psi\bar\psi}$), and we have made use of $Y_\psi\equiv Y_{\bar\psi}$ due to CP conservation. For simplicity, we assume $M_\phi$, $M_{Z'}\gtrsim M_{\nu_R}$ so that the $B-L$ scalar $\phi$ and gauge boson $Z'$ do not participate in the Boltzmann equations explicitly, but $Z'$ contribute to $\gamma_{\nu_R\nu_R\to f\bar f}$ and $\gamma_{\psi\bar\psi\to f\bar f}$ (where $f$ denotes the SM fermions in equilibrium) via the off-shell $s$-channel diagrams. In principle, the decay $\psi\to jjj$ and scattering $\psi j\to jj$ induced by the dimension-6 operator in \Eq{L6} also affect the evolution of $Y_\psi$, but the effect is completely negligible due to the large $\Lambda$. Instead, the late time $\psi$ decay after $\nu_R$ freeze-out plays an important role, as discussed in the next subsection.

\subsection{Decay of $\psi$ and the dark matter relic abundance}

Due to its long life time, $\psi$ can be treated as a stable particle during the freeze-out of $\nu_R$, and its ``relic abundance'' is given by $Y_\psi^\infty=Y_\psi(z\to\infty)$. After freeze-out, the energy density of $\psi$ scales as $a^{-3}$ where $a(t)$ is the Friedmann-Lemaitre-Robertson-Walker (FLRW) scale factor, while the radiation energy density scales as $a^{-4}$. As a result, if $\psi$ is sufficiently long-lived, it dominates the Universe energy at the late time, and its decay would generate significant entropy that further suppresses the $\nu_R$ relic abundance. The dilution factor can be estimated quickly by the following considerations: $\psi$ decays at $T=T_\psi$ when $\Gamma_\psi\sim H$, i.e.
\be
\Gamma_\psi^2\sim H^2|_{T=T_\psi}\approx\frac{8\pi}{3M_{\rm Pl}^2}M_\psi Y_\psi^\infty\left(\frac{2\pi^2}{45}g_*T_\psi^3\right),
\ee
If $\psi$ decays to $jjj$ very promptly at $t_\psi=1/\Gamma_\psi$ and reheats the Universe up to $T'_\psi$, by energy conservation we have
\be
\frac{\pi^2}{30}g_*T_\psi'^4=M_\psi Y_\psi ^\infty\left(\frac{2\pi^2}{45}g_*T_\psi^3\right).
\ee
Combining these two equations, one obtains the entropy enhancement factor (or equivalently, the DM density dilution factor)
\be
\Delta_\psi=\frac{S_{\rm after}}{S_{\rm before}}\approx\left(\frac{T_\psi'}{T_\psi}\right)^3\sim g_*^{1/4}\frac{M_\psi Y_\psi^\infty}{\sqrt{M_{\rm Pl}\Gamma_\psi}},
\ee
and the final DM relic abundance is $Y_{\rm DM}=Y_{\nu_R}^\infty/\Delta_\psi$. A more detailed treatment in Ref.~\cite{kolb1981early} yields
\be
\Delta_\psi\approx1.83\,\big\langle g_*^{1/3}\big\rangle^{3/4}\frac{M_\psi Y_\psi^\infty}{\sqrt{M_{\rm Pl}\Gamma_\psi}},
\ee
which will be adopted in our numerical study.

\begin{table}\footnotesize\renewcommand\arraystretch{1.5}\centering
\begin{tabular}{c|c|c|c||c|c|c|c}\hline\hline
 & $M_{\nu_R}$ & $\lambda_1$ & $\Lambda$ & $T_\psi$ & $T'_\psi$ & $\Delta_\psi$ & $\tau_{\nu_R}$ \\ \hline
 BP1 & $1.5\times10^{13}$ GeV & 1.5 & $2\times10^{18}$ GeV & 3.0 GeV & $2.0\times10^3$ GeV & $3.1\times10^{8}$ & $1.0\times10^{28}$ s \\ \hline
 BP2 & $1.0\times10^{12}$ GeV & 0.74 & $1\times10^{17}$ GeV & 2.7 GeV & $9.4\times10^2$ GeV & $4.1\times10^7$ & $2.3\times10^{29}$ s \\ \hline
 BP3 & $4.0\times10^{10}$ GeV & 0.40 & $2\times10^{15}$ GeV & 7.7 GeV & $7.5\times10^2$ GeV & $9.1\times10^5$ & $2.5\times10^{29}$ s \\ \hline
 BP4 & $3.0\times10^{9}$ GeV & 0.30 & $1\times10^{14}$ GeV & 16 GeV & $4.6\times10^2$ GeV & $2.4\times10^4$ & $4.1\times10^{29}$ s\\ \hline\hline
\end{tabular} 
\caption{The BPs for the superheavy RHN DM scenario. $g_{B-L}=1$, $\lambda_R=0.1$, $M_{\nu_R}=1.9M_\psi$, $M_S=2M_{\nu_R}$ and $\lambda_1=\lambda_2$ are fixed, and the first three columns $M_{\nu_R}$ (RHN mass), $\lambda_1$ ($\nu_R$-$\psi$-$S$ coupling) and $\Lambda$ ($\Z_2$ breaking scale) are free parameters. The $\psi$ decay temperature $T_\psi$ and reheating temperature $T_\psi'$, dilution factor $\Delta_\psi$ and the DM life time $\tau_{\nu_R}$ are derived, see the text.}\label{tab:BPs}
\end{table}

\begin{figure}
\centering
\subfigure{
\includegraphics[scale=0.45]{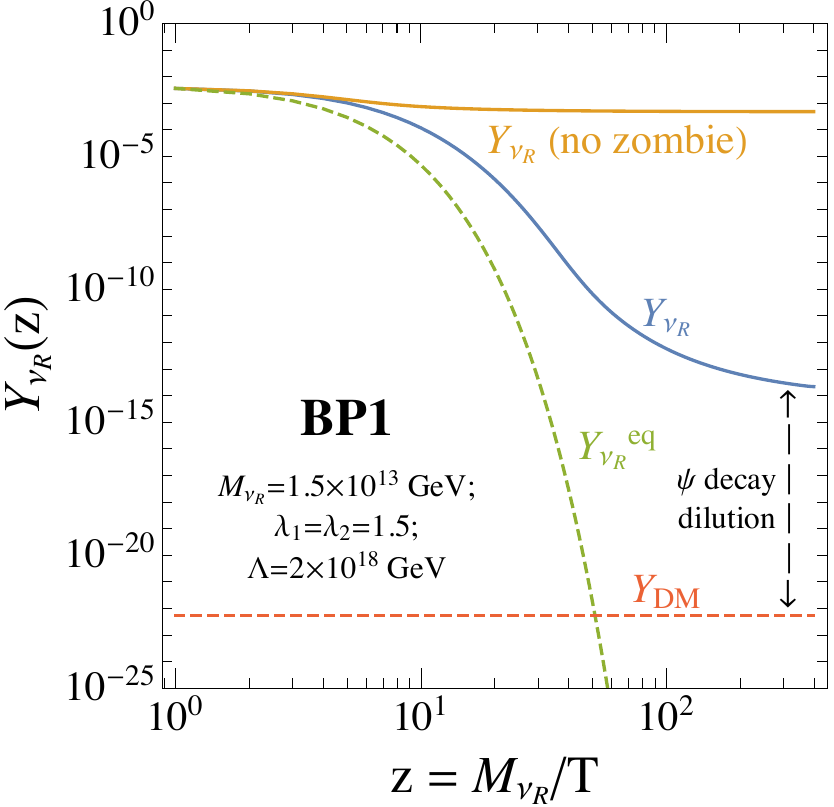}}\qquad
\subfigure{
\includegraphics[scale=0.45]{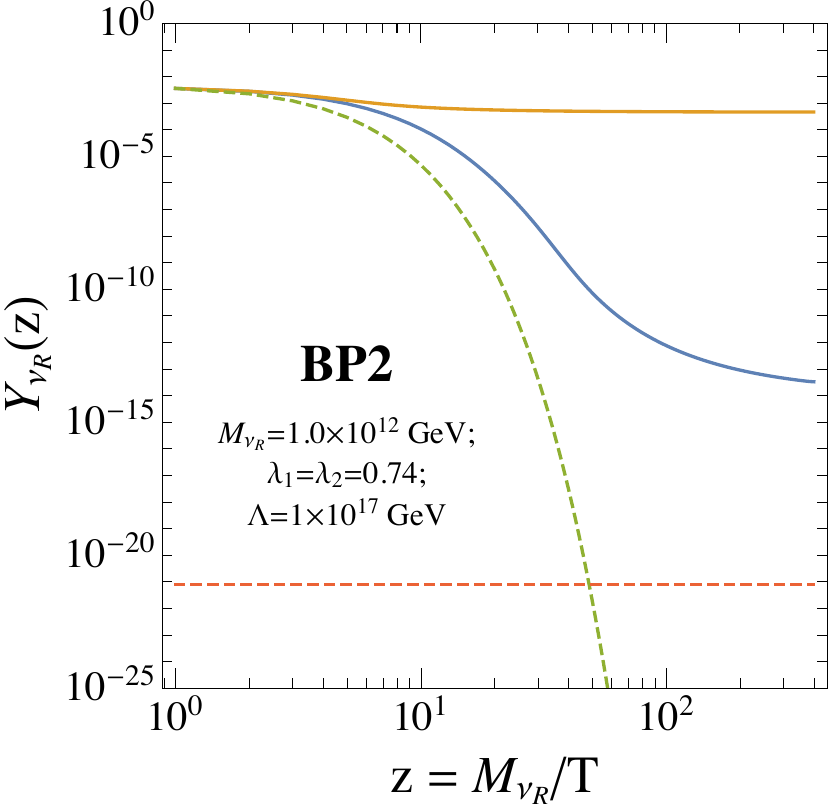}}
\subfigure{
\includegraphics[scale=0.45]{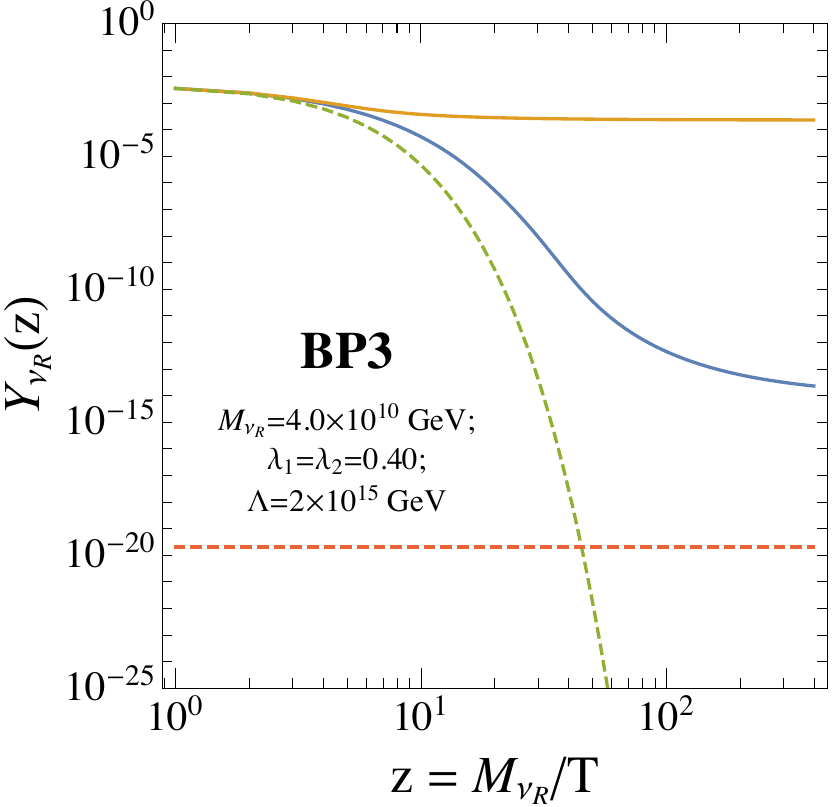}}\quad
\subfigure{
\includegraphics[scale=0.45]{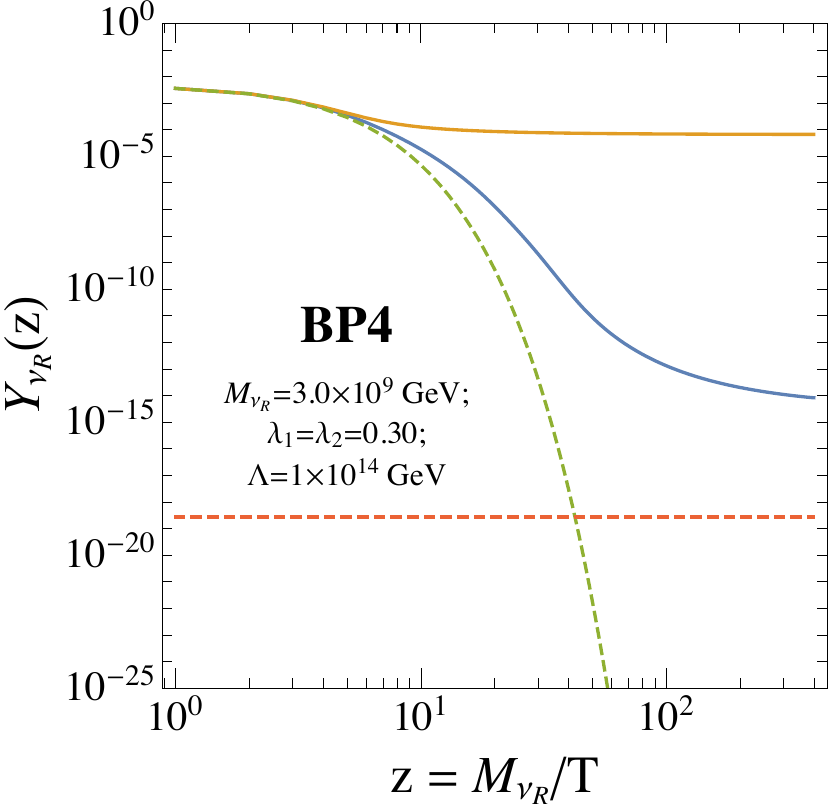}}
\caption{Evolution of the $\nu_R$ abundance for the four BPs in Table~\ref{tab:BPs}. $Y_{\nu_R}$ and $Y_{\nu_R}^{\rm eq}$ are shown in blue and green curves, respectively. For comparison, we also show in orange curves the $Y_{\nu_R}$ evolution without the zombie collision.}
\label{fig:BPs}
\end{figure}

Now we are ready to calculate the DM relic abundance by solving the Boltzmann equations (\ref{Boltzmann_1}) and (\ref{Boltzmann_2}) to get $Y_{\nu_R}^\infty$, and combining the dilution factor $\Delta_\psi$. For the numerical study, we fix $g_{B-L}=1$, $\lambda_R=0.1$, $M_{\nu_R}=1.9M_\psi$, $M_S=2M_{\nu_R}$ and $\lambda_1=\lambda_2$, and vary $M_{\nu_R}$, $\lambda_1$ and $\Lambda$ to find four benchmark points (BPs) listed in Table~\ref{tab:BPs}. All BPs can yield the correct relic abundance
\be
\Omega_{\rm DM}h^2=\frac{8\pi h^2}{3M_{\rm Pl}^2H_0^2}\frac{Y_{\nu_R}^\infty s_0}{\Delta_\psi}M_{\nu_R}\approx0.12,
\ee
where $H_0=100\,h~{\rm km/s/Mpc}$ (with $h=0.674$) and $s_0=2891.2~{\rm cm}^{-3}$ are current Hubble constant and entropy density, respectively~\cite{ParticleDataGroup:2020ssz}. As we can see, due to the enhanced cross section from the light sterile neutrino (i.e. zombie collision) and the dilution factor from $\psi$ decay, a $1.5\times10^{13}$ GeV RHN can still yield the correct DM abundance with a moderate coupling $\lambda_1=\lambda_2=1.5$. Even for a small coupling $\lambda_1=\lambda_2=0.30$, the RHN DM can be as heavy as $3.0\times10^9$ GeV, well above the GK bound. In our mass setup, $\nu_R$ can decay to 9 jets via 2 off-shell $\psi$'s, and the life time is calculated by the {\tt FeynRules}~\cite{Alloul:2013bka} and {\tt MadGraph5\_aMC@NLO}~\cite{Alwall:2014hca} packages. $\tau_{\nu_R}\gtrsim10^{27}$ s is satisfied in all BPs, as required by the diffuse gamma-ray spectrum~\cite{Cirelli:2012ut,Essig:2013goa,Blanco:2018esa}.

To distinguish and compare the contributions from ``zombie collision'' and ``$\psi$ decay dilution'' to the DM relic density, we plot the $Y_{\nu_R}$ evolution as blue curves in the BPs in Fig.~\ref{fig:BPs}. The equilibrium distribution $Y_{\nu_R}^{\rm eq}$ is shown in green curves for reference, such that one can clearly see the $\nu_R$ deviates from the thermal equilibrium and freeze-out to a constant abundance $Y_{\nu_R}^\infty$. The late time decay of $\psi$ will dilute the freeze-out abundance to $Y_{\nu_R}^\infty/\Delta_\psi$, which is eventually $Y_{\rm DM}\approx0.8~{\rm eV}/M_{\nu_R}$, as shown in red dashed straight lines. To manifest the importance of the zombie collision, we also plot the $Y_{\rm \nu_R}$ evolution assuming $\lambda_1=\lambda_2=0$ in orange curves. In that case, $\nu_R$ freeze-out in a very high abundance and hence cannot provide the correct DM density even with the help of the $\psi$ decay dilution. By comparing the curves of $Y_{\nu_R}$, $Y_{\nu_R}$ (without zombie) and $Y_{\rm DM}$, one can see that the contributions from zombie collision and $\psi$ decay dilution are comparable.

\section{Cosmic strings and the gravitational wave signals}\label{sec:gws}

The BPs obtained in the last section are for $M_{\nu_R}\gtrsim10^9$ GeV, much higher than the available energy scale of any traditional DM direct, indirect or collider searches, making it almost hopeless to test superheavy DM scenario. However, the recently developed GW astronomy offers an unique opportunity to probe this scenario: as the RHN mass $M_{\nu_R}$ is associated with a high-scale $U(1)_{B-L}$ breaking, which forms cosmic strings that can generate detectable stochastic GW signals~\cite{Buchmuller:2013lra,Dror:2019syi,Auclair:2019wcv,Fornal:2020esl,Samanta:2020cdk,Masoud:2021prr,Buchmuller:2021mbb}.

Cosmic strings are one dimensional topological defects form during a spontaneous symmetry breaking if the topology of the vacuum is not simply connected~\cite{Nielsen:1973cs}. In our scenario, we consider Nambu-Goto cosmic strings that can form after the $U(1)_{B-L}$ breaking, and the energy density per unit length $\mu\sim v_\phi^2$. A very important observable for the cosmic strings is the dimensionless combination
\be\label{Gmu}
G\mu\sim \frac{v_\phi^2}{M_{\rm Pl}^2}\sim10^{-10}\times\left(\frac{M_{\nu_R}/\lambda_R}{10^{14}~{\rm GeV}}\right)^2,
\ee
where $G$ is the Newton's constant of gravitation. After formation, the collisions and self-interactions of strings produce sub-horizon, non-self-interacting string loops, which emit GWs via cusp, kink and kink-kink collisions, and they produce GWs throughout the Universe history. The incoherent superposition of such continuous emission results in today's stochastic GW signals.

According to above physical picture, the GW spectrum today can be expressed as
\be\label{CS_GWs}
\Omega_{\rm GW}(f)h^2=\frac{8\pi h^2}{3M_{\rm Pl}^2H_0^2}\int_0^{t_0}dt\left(\frac{a(t)}{a(t_0)}\right)^3\int_0^\infty d\ell\,n_{\rm CS}(\ell,t)P_{\rm GW}\left(\frac{a(t_0)}{a(t)}f,\ell\right),
\ee
where $a(t)$ is the FLRW scale factor and $t_0$ is the current cosmic time. $n_{\rm CS}(\ell,t)$ is the number density of sub-horizon string loops with invariant length $\ell$ at cosmic time $t$, while $P_{\rm GW}(f,\ell)$ is the loop power spectrum describing the power of GW with frequency $f$ emitted from a loop with length $\ell$. We follow the method described in Ref.~\cite{Auclair:2019wcv} to calculate the GWs from the Nambu-Goto string~\cite{Vachaspati:1984gt} by transforming \Eq{CS_GWs} into~\cite{Auclair:2019wcv,Blanco-Pillado:2017oxo}
\be
\Omega_{\rm GW}(f)h^2=\frac{8\pi h^2}{3M_{\rm Pl}^2H_0^2}G\mu^2 f\sum_{n=1}^\infty C_n(f)P_n,
\ee
where $n=1$, 2, ..., labels the radiation frequencies $\omega_n=2\pi n/(\ell/2)$, and $P_n$ is the corresponding average loop power spectrum which we use the numerical results from Ref.~\cite{Blanco-Pillado:2017oxo}, and
\be\label{Cn}
C_n=\frac{2n}{f^2}\int_0^{\infty}\frac{dz}{H(z)(1+z)^6}n_{\rm CS}\left(\frac{2n}{(1+z)f},t(z)\right),
\ee
where we have transformed the integral variable from time $t$ to the redshift $z$.

To evaluate the integration in \Eq{Cn}, the cosmic time and Hubble constant should be expressed as functions of redshift
\be
t(z)=\int_z^\infty\frac{dz'}{H(z')(1+z')};\quad H(z)=H_0\sqrt{\Omega_r\mG(z)(1+z)^4+\Omega_m(1+z)^3+\Omega_\Lambda},
\ee
with the abundance~\cite{Planck:2018vyg}
\be
\Omega_r=9.1476\times10^{-5},\quad \Omega_m=0.308,\quad\Omega_\Lambda=1-\Omega_r-\Omega_m,
\ee
and the function
\be
\mG(z)=\frac{g_*(t)}{g_*(t_0)}\left(\frac{g_S(t_0)}{g_S(t)}\right)^{4/3}\approx\begin{cases}~1,&z<10^9;\\~0.83,&10^9<z<2\times10^{12};\\~0.39,&z>2\times10^{12},\end{cases}
\ee
enfolds the change of relativistic degrees of freedom at $e^+e^-$ annihilation (200 keV, $z=10^9$) and QCD phase transition (200 MeV, $z=2\times10^{12}$)~\cite{Binetruy:2012ze}. The cosmic string number density is~\cite{Blanco-Pillado:2013qja}
\be
n_{\rm CS}^r(\ell,t)=\frac{0.18}{t^{3/2}(\ell+\Gamma G\mu t)^{5/2}},\quad(\ell\leqslant0.1\,t);
\ee
for the loops in radiation dominated era, and
\be
n_{\rm CS}^{r,m}(\ell,t)=\frac{0.18\, t_{\rm eq}^{1/2}}{t^{2}(\ell+\Gamma G\mu t)^{5/2}},\quad (\ell\leqslant0.09\, t_{\rm eq}-\Gamma G\mu t);
\ee
for the loops produced in radiation dominated era but survive until matter domination, where $\Gamma=50$~\cite{Blanco-Pillado:2017oxo}, and $t_{\rm eq}=2.25\times10^{36}~{\rm GeV}^{-1}$ is the matter-radiation equality time. In the matter dominated era, the loop number density is
\be
n_{\rm CS}^m(\ell,t)=\frac{0.27-0.45(\ell/t)^{0.31}}{t^2(\ell+\Gamma G\mu t)^2},\quad (\ell\leqslant0.18\, t).
\ee
Up to now, the GW spectrum is calculable for a given $G\mu$.

\begin{figure}
\centering
\subfigure{
\includegraphics[scale=0.6]{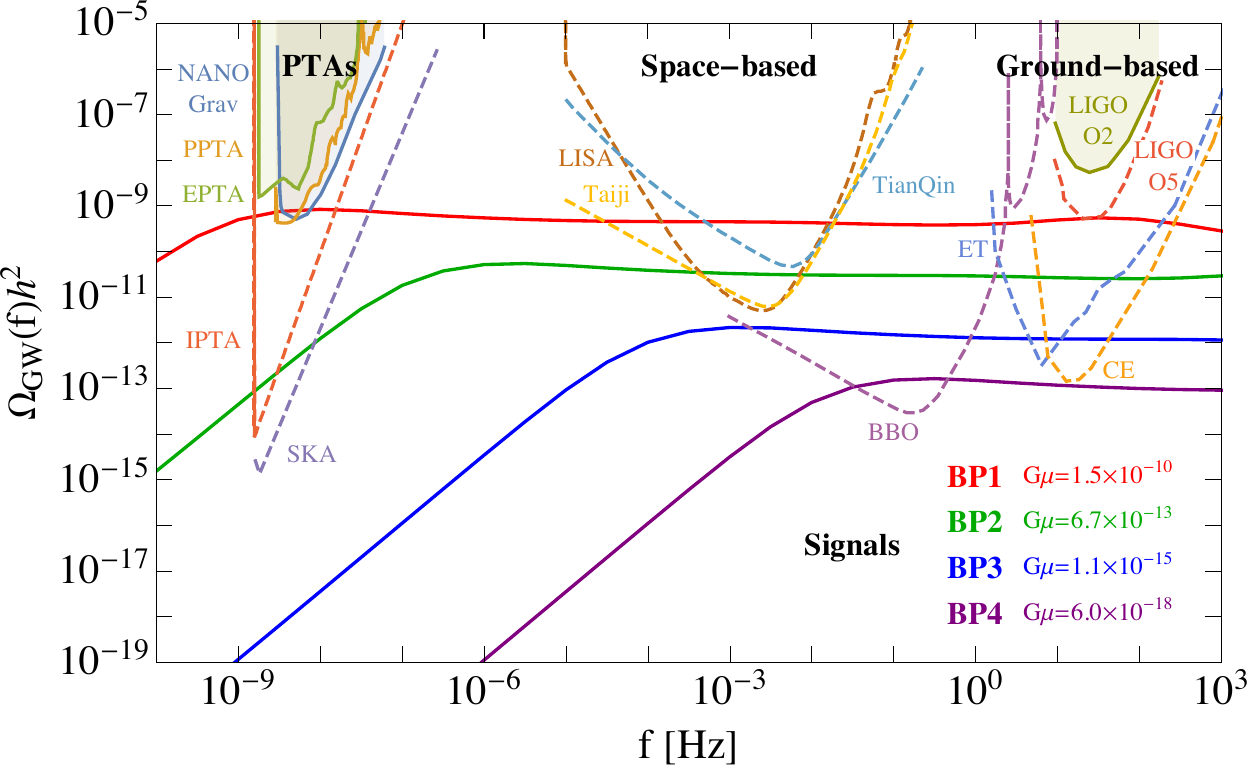}}
\caption{The GW signals from the BPs in Table~\ref{tab:BPs}, and the relevant sensitivity curves of current or future GW detectors. The existing constraints and projected sensitivities are shown as shaded regions and dashed lines, respectively.}
\label{fig:GWs}
\end{figure}

Using \Eq{Gmu}, we evaluate the GW signals for the four BPs in Table~\ref{tab:BPs}, and plot them as red, green, blue and purple  lines in Fig.~\ref{fig:GWs}. Due to the continuous GW emission, the signal spectra are flat in a large frequency range. The sensitivity curves of current and future GW detectors are also plotted in the figure, including the pulsar timing arrays (PTAs) NANOGrav~\cite{McLaughlin:2013ira,NANOGRAV:2018hou,Aggarwal:2018mgp,Brazier:2019mmu}, PPTA~\cite{Manchester:2012za,Shannon:2015ect}, EPTA~\cite{Kramer:2013kea,Lentati:2015qwp,Babak:2015lua}, IPTA~\cite{Hobbs:2009yy,Manchester:2013ndt,Verbiest:2016vem,Hazboun:2018wpv} and SKA~\cite{Carilli:2004nx,Janssen:2014dka,Weltman:2018zrl}, the space-based laser interferometers LISA~\cite{LISA:2017pwj}, BBO~\cite{Crowder:2005nr}, TianQin~\cite{TianQin:2015yph,Hu:2017yoc,TianQin:2020hid} and Taiji~\cite{Hu:2017mde,Ruan:2018tsw}, and the ground-based interferometers LIGO~\cite{LIGOScientific:2014qfs,LIGOScientific:2019vic}, CE~\cite{Reitze:2019iox} and ET~\cite{Punturo:2010zz,Hild:2010id,Sathyaprakash:2012jk}.\footnote{The sensitivity curves shown here are strain noise spectra. For the corresponding power-law-integrated sensitivity curves and peak-integrated sensitivity curves, see Ref.~\cite{Schmitz:2020syl} and the references therein.} The existing constraints are shown as shaded regions, while the projected sensitivities are plotted as dashed lines. 

The current searches for the SGWB of PTAs constrains $G\mu<10^{-11}$~\cite{Ringeval:2017eww,Blanco-Pillado:2017rnf} which requires the $M_{\nu_R}/\lambda_R<3.2\times10^{13}$ GeV according to \Eq{Gmu}.  
In Fig.~\ref{fig:GWs}, we can see that BP1 (with $M_{\nu_R}=1.5\times10^{13}$ GeV) is already within the reach of the PPTA and NANOGrav experiments, and the null results of these observations suggest that the BP1 is excluded; but a DM with mass slightly lower than BP1 is still allowed and can be tested in the near future. Interestingly, recently the NANOGrav collaboration reported a common-spectrum process based on the 12.5-yr data set~\cite{NANOGrav:2020bcs}, which might be a hint for the stochastic GW background from cosmic string~\cite{Ellis:2020ena,Bian:2020urb,Blasi:2020mfx}. However, there is some discrepancy between the results of the NANOGrav and PPTA collaborations, and we still have to wait for the future data of the PTA experiments or even the crosscheck from the space- and ground-based detectors to reveal the origin of the excess. 
The GW signals from BP2 and BP3 can be accessed by a few future detectors, while the GWs from BP4 (with $M_{\nu_R}=3.0\times10^9$ GeV) can be reached only by BBO and CE. Therefore, the cosmic strings induced GWs can probe our scenario for RHN DM mass between $\sim10^9$ GeV and $\sim10^{13}$ GeV. We also note that there are still considerable uncertainties in calculating the cosmic string GWs, and the expected reach for DM mass range might vary when more accurate treatments are available.\footnote{For example, the LIGO-Virgo O3 data can exclude $G\mu\gtrsim 4\times 10^{-15}$ for some specific string network models~\cite{LIGOScientific:2021nrg}, much stronger than the results presented in Fig.~\ref{fig:GWs}.}

\section{Conclusion}\label{sec:conclusion}

In this article, we have realized the zombie collision DM mechanism in an extended $U(1)_{B-L}$ model, showing it allows RHN DM mass up to $10^{13}$ GeV with a moderate interaction coupling. Such a superheavy DM scenario benefits from both the exponentially annihilation rate due to the lighter sterile neutrino, and the late time entropy production from the sterile neutrino decay.

As the DM mass significantly exceeds the sensitivity regions of the traditional DM detection experiments, we propose to probe the zombie collision DM scenario via GW astronomy by detecting the GW signals from the cosmic strings formed when the $U(1)_{B-L}$ breaking. Calculations show that RHN DM mass between $\mO(10^9~{\rm GeV})$ and $\mO(10^{13}~{\rm GeV})$ can be probed by future GW experiments; especially, for DM with mass $\sim10^{13}$ GeV, the signal might already be reached by the recent NANOGrav results, but more data are needed to clarify this.

\acknowledgments

We would like to thank Xucheng Gan, Huai-Ke Guo, Yi-Lei Tang, Daniele Teresi, Shao-Jiang Wang and Bin Zhu for useful discussions. Ligong Bian was supported by the National Natural Science Foundation of China under the grants Nos.12075041, 12047564, and the Fundamental Research Funds for the Central Universities of China (No. 2021CDJQY-011 and No. 2020CDJQY-Z003),  and Chongqing Natural Science Foundation (Grants No.cstc2020jcyj-msxmX0814). Xuewen Liu was supported by the National Natural Science Foundation of China under the Grants No. 12005180, and by the Natural Science Foundation of Shandong Province under the Grant No. ZR2020QA083. KPX is supported by Grant Korea NRF-2019R1C1C1010050.

\appendix
\section{The interaction rates}\label{app:rates}

The interaction rates in Section~\ref{sec:freeze-out} are defined as
\be\label{gamma}\begin{split}
\gamma_{ab\to cd}\equiv\ave{\sigma_{ab\to cd}v}n_a^{\rm eq}n_b^{\rm eq}=&~\frac{M_{\nu_R}}{64\pi^4z}\int_{s_{\rm min}}^\infty ds\,\hat\sigma_{ab\to cd}(s)\sqrt{s}K_1\left(\frac{\sqrt{s}}{M_{\nu_R}}z\right)\\
=&~\frac{M_{\nu_R}^4}{64\pi^4z}\int_{x_{\rm min}}^\infty dx\,\hat\sigma_{ab\to cd}(M_{\nu_R}^2x)\sqrt{x}K_1(z\sqrt{x}),
\end{split}\ee
where $K_1$ is the modified Bessel function of the first kind, $s_{\rm min}=\max\{(M_a+M_b)^2,~(M_c+M_d)^2\}$, $x=s/M_{\nu_R}^2$, and the reduced cross section is
\be
\hat\sigma_{ab\to cd}(s)\equiv 2\,\sigma_{ab\to cd}(s)\cdot s\cdot\left[1-2\left(\frac{M_a^2}{s}+\frac{M_b^2}{s}\right)+\left(\frac{M_a^2}{s}-\frac{M_b^2}{s}\right)^2\right],
\ee
where $\sigma_{ab\to cd}$ is the Lorentz invariant cross section, which is a function of $s$. All initial and final state spin and internal degrees of freedom have been summed up. Our convention of $\gamma_{ab\to cd}$ is the same as Ref.~\cite{Giudice:2003jh}.

Since the explicit expressions for the interaction rates in Section~\ref{sec:freeze-out} are too tedious to be shown in the article, below we only list the amplitude squares for the corresponding processes in our mechanism. With the amplitude squares in hand, one can easily derives the cross sections and interaction rates. Assume the collision to be $a(p_1)b(p_2)\to c(p_3)d(p_4)$, and define the Mandelstam variables as $s=(p_1+p_2)^2$ and $t=(p_3-p_1)^2$, for the zombie collisions we have
\begin{multline}
\sum_{\rm spins}|i\mM_{\nu_R\psi\to\psi\psi}|^2=\lambda_1^2\lambda_2^2\left(M_S^2-t\right)^{-2}\left(s+t-3 M_\psi^2+M_S^2-M_{\nu_R}^2\right)^{-2}\times\\
\Big\{\left(M_S^2-t\right) \left(2M_\psi^2-s-t\right) \left(M_\psi^2-M_{\nu_R}^2+s+t\right) 
\left(s+3t-3M_\psi^2-M_S^2-M_{\nu_R}^2\right)\\
+\Big[\left(4 M_\psi^2-t\right) \left(M_\psi^2+M_{\nu_R}^2-t\right) \left(2s+3t-6M_\psi^2+M_S^2-2M_{\nu_R}^2\right)\\ 
+\left(4 M_\psi^2-s\right) \left(M_S^2-t\right)\left(M_\psi^2+M_{\nu_R}^2-s\right)\Big]
\left(s+t-3M_\psi^2+M_S^2-M_{\nu_R}^2\right)\Big\},
\end{multline}
and
\begin{multline}
\sum_{\rm spins}|i\mM_{\nu_R\psi\to\psi\bar\psi}|^2=\lambda_1^2\lambda_2^2 \left(M_S^2-t\right)^{-2}\left[\left(s-M_S^2\right)^2+M_S^2\Gamma_S^2\right]^{-1}\times \\
\Big\{\left(4 M_\psi^2-t\right)
   \left(M_\psi^2+M_{\nu_R}^2-t\right) \left(2 \Gamma_S^2
   M_S^2+\left(M_S^2-s\right) \left(M_S^2-2 s+t\right)\right)\\
   +\left(t-M_S^2\right)
   \Big[\left(s-4M_\psi^2\right) \left(M_\psi^2+M_{\nu_R}^2-s\right)
   \left(M_S^2+s-2 t\right)\\
   +\left(s-M_S^2\right) \left(s+t-2M_\psi^2\right)
   \left(s+t+M_\psi^2-M_{\nu_R}^2\right)\Big]\Big\}.
\end{multline}
Note that in the process $\nu_R\psi\to\psi\bar\psi$, the mediator $S$ can be in the $s$-channel, thus its width should be included to make the integral in \Eq{gamma} converge.

For the pair annihilation of $\nu_R\nu_R\to\psi\bar\psi$, there are both contributions from the Yukawa interactions and $U(1)_{B-L}$ gauge interactions,
\begin{multline}
\sum_{\rm spins}|i\mM_{\nu_R\nu_R\to\psi\bar\psi}|^2=\lambda_1^4 \Big\{\frac{2 M_{\nu_R}^2(s-2 M_\psi^2)}{(t-M_S^2)(s+t-2 M_\psi^2+M_S^2-2 M_{\nu_R}^2)}\\
+\frac{(M_\psi^2+M_{\nu_R}^2-s-t)^2}{(s+t-2 M_\psi^2+M_S^2-2M_{\nu_R}^2)^2}+\frac{(M_\psi^2+M_{\nu_R}^2-t)^2}{(M_S^2-t)^2}\Big\}\\
+8g_{B-L}^4\frac{2 M_\psi^4-4 M_\psi^2(M_{\nu_R}^2+t)+2 M_{\nu_R}^4-4
   M_{\nu_R}^2 (s+t)+s^2+2 s t+2 t^2}{(s-M_{Z'}^2)^2+M_{Z'}^2\Gamma_{Z'}^2},
\end{multline}
while for $\nu_R\nu_R\to\psi\psi$, only Yukawa interactions contribute,
\begin{multline}
\sum_{\rm spins}|i\mM_{\nu_R\nu_R\to\psi\psi}|^2=\lambda_1^4\left(M_S^2-t\right)^{-2}\left(s+t-2 M_\psi^2+M_S^2-2 M_{\nu_R}^2\right)^{-2}\times \\
\Big\{\left(s+2t-2M_\psi^2-2 M_{\nu_R}^2\right)^2\left[t \left(s-2
   \left(M_\psi^2+M_{\nu_R}^2\right)\right)+\left(M_\psi^2+M_{\nu_R}^2\right)^2-sM_S^2+t^2\right]\\
   +\left(s-2 M_\psi^2\right) \left(M_S^2-t\right) \left(s-2 M_{\nu_R}^2\right)
   \left(s+t-2 M_\psi^2+M_S^2-2 M_{\nu_R}^2\right)\Big\}.
\end{multline}

Finally, the $Z'$ mediated annihilation to the SM fermions reads
\be\begin{split}
\sum_{\rm spins}|i\mM_{\nu_R\nu_R\to f\bar f}|^2=&~\frac{13}{2}\times 8g_{B-L}^4\frac{(s+t-2M_{\nu_R}^2)^2+t^2-2M_{\nu_R}^4}{(s-M_{Z'}^2)^2+M_{Z'}^2\Gamma_{Z'}^2},\\
\sum_{\rm spins}|i\mM_{\psi\bar\psi\to f\bar f}|^2=&~\frac{13}{2}\times 8g_{B-L}^4\frac{(s+t)^2+(2M_\psi^2-t)^2-2M_\psi^4}{(s-M_{Z'}^2)^2+M_{Z'}^2\Gamma_{Z'}^2}.
\end{split}\ee
Here the factor $13/2$ comes from the summation of SM fermionic degrees of freedom.

\bibliographystyle{JHEP-2-2.bst}
\bibliography{references}

\providecommand{\href}[2]{#2}\begingroup\raggedright\begin{thebibliography}{100}

\bibitem{Lee:1977ua}
B.~W. Lee and S.~Weinberg, ``{Cosmological Lower Bound on Heavy Neutrino
  Masses},''\href{http://dx.doi.org/10.1103/PhysRevLett.39.165}{\emph{Phys.
  Rev. Lett.} {\bf 39} (1977) 165--168}.

\bibitem{Kolb:1990vq}
E.~W. Kolb and M.~S. Turner, \emph{{The Early Universe}}, vol.~69.
\newblock 1990.

\bibitem{Bertone:2004pz}
G.~Bertone, D.~Hooper and J.~Silk, ``{Particle dark matter: Evidence,
  candidates and
  constraints},''\href{http://dx.doi.org/10.1016/j.physrep.2004.08.031}{\emph{Phys.
  Rept.} {\bf 405} (2005) 279--390},
  [\href{https://arxiv.org/abs/hep-ph/0404175}{{\tt hep-ph/0404175}}].

\bibitem{Lisanti:2016jxe}
M.~Lisanti, ``{Lectures on Dark Matter Physics},'' in \emph{{Theoretical
  Advanced Study Institute in Elementary Particle Physics}: {New Frontiers in
  Fields and Strings}}, 3, 2016.
\newblock \href{https://arxiv.org/abs/1603.03797}{{\tt 1603.03797}}.
\newblock \href{http://dx.doi.org/10.1142/9789813149441_0007}{DOI}.

\bibitem{Planck:2018vyg}
{\scshape Planck} collaboration, N.~Aghanim et~al., ``{Planck 2018 results. VI.
  Cosmological
  parameters},''\href{http://dx.doi.org/10.1051/0004-6361/201833910}{\emph{Astron.
  Astrophys.} {\bf 641} (2020) A6},
  [\href{https://arxiv.org/abs/1807.06209}{{\tt 1807.06209}}].

\bibitem{ParticleDataGroup:2020ssz}
{\scshape Particle Data Group} collaboration, P.~A. Zyla et~al., ``{Review of
  Particle Physics},''\href{http://dx.doi.org/10.1093/ptep/ptaa104}{\emph{PTEP}
  {\bf 2020} (2020) 083C01}.

\bibitem{Schumann:2019eaa}
M.~Schumann, ``{Direct Detection of WIMP Dark Matter: Concepts and
  Status},''\href{http://dx.doi.org/10.1088/1361-6471/ab2ea5}{\emph{J. Phys. G}
  {\bf 46} (2019) 103003}, [\href{https://arxiv.org/abs/1903.03026}{{\tt
  1903.03026}}].

\bibitem{Gaskins:2016cha}
J.~M. Gaskins, ``{A review of indirect searches for particle dark
  matter},''\href{http://dx.doi.org/10.1080/00107514.2016.1175160}{\emph{Contemp.
  Phys.} {\bf 57} (2016) 496--525},
  [\href{https://arxiv.org/abs/1604.00014}{{\tt 1604.00014}}].

\bibitem{Kitano:2010fa}
R.~Kitano, H.~Ooguri and Y.~Ookouchi, ``{Supersymmetry Breaking and Gauge
  Mediation},''\href{http://dx.doi.org/10.1146/annurev.nucl.012809.104540}{\emph{Ann.
  Rev. Nucl. Part. Sci.} {\bf 60} (2010) 491--511},
  [\href{https://arxiv.org/abs/1001.4535}{{\tt 1001.4535}}].

\bibitem{Griest:1989wd}
K.~Griest and M.~Kamionkowski, ``{Unitarity Limits on the Mass and Radius of
  Dark Matter
  Particles},''\href{http://dx.doi.org/10.1103/PhysRevLett.64.615}{\emph{Phys.
  Rev. Lett.} {\bf 64} (1990) 615}.

\bibitem{Kolb:1998ki}
E.~W. Kolb, D.~J.~H. Chung and A.~Riotto,
  ``{WIMPzillas!},''\href{http://dx.doi.org/10.1063/1.59655}{\emph{AIP Conf.
  Proc.} {\bf 484} (1999) 91--105},
  [\href{https://arxiv.org/abs/hep-ph/9810361}{{\tt hep-ph/9810361}}].

\bibitem{Hui:1998dc}
L.~Hui and E.~D. Stewart, ``{Superheavy dark matter from thermal
  inflation},''\href{http://dx.doi.org/10.1103/PhysRevD.60.023518}{\emph{Phys.
  Rev. D} {\bf 60} (1999) 023518},
  [\href{https://arxiv.org/abs/hep-ph/9812345}{{\tt hep-ph/9812345}}].

\bibitem{Chung:1998rq}
D.~J.~H. Chung, E.~W. Kolb and A.~Riotto, ``{Production of massive particles
  during
  reheating},''\href{http://dx.doi.org/10.1103/PhysRevD.60.063504}{\emph{Phys.
  Rev. D} {\bf 60} (1999) 063504},
  [\href{https://arxiv.org/abs/hep-ph/9809453}{{\tt hep-ph/9809453}}].

\bibitem{Chung:2001cb}
D.~J.~H. Chung, P.~Crotty, E.~W. Kolb and A.~Riotto, ``{On the Gravitational
  Production of Superheavy Dark
  Matter},''\href{http://dx.doi.org/10.1103/PhysRevD.64.043503}{\emph{Phys.
  Rev. D} {\bf 64} (2001) 043503},
  [\href{https://arxiv.org/abs/hep-ph/0104100}{{\tt hep-ph/0104100}}].

\bibitem{Harigaya:2014waa}
K.~Harigaya, M.~Kawasaki, K.~Mukaida and M.~Yamada, ``{Dark Matter Production
  in Late Time
  Reheating},''\href{http://dx.doi.org/10.1103/PhysRevD.89.083532}{\emph{Phys.
  Rev. D} {\bf 89} (2014) 083532}, [\href{https://arxiv.org/abs/1402.2846}{{\tt
  1402.2846}}].

\bibitem{Davoudiasl:2015vba}
H.~Davoudiasl, D.~Hooper and S.~D. McDermott, ``{Inflatable Dark
  Matter},''\href{http://dx.doi.org/10.1103/PhysRevLett.116.031303}{\emph{Phys.
  Rev. Lett.} {\bf 116} (2016) 031303},
  [\href{https://arxiv.org/abs/1507.08660}{{\tt 1507.08660}}].

\bibitem{Randall:2015xza}
L.~Randall, J.~Scholtz and J.~Unwin, ``{Flooded Dark Matter and S Level
  Rise},''\href{http://dx.doi.org/10.1007/JHEP03(2016)011}{\emph{JHEP} {\bf 03}
  (2016) 011}, [\href{https://arxiv.org/abs/1509.08477}{{\tt 1509.08477}}].

\bibitem{Harigaya:2016vda}
K.~Harigaya, T.~Lin and H.~K. Lou, ``{GUTzilla Dark
  Matter},''\href{http://dx.doi.org/10.1007/JHEP09(2016)014}{\emph{JHEP} {\bf
  09} (2016) 014}, [\href{https://arxiv.org/abs/1606.00923}{{\tt 1606.00923}}].

\bibitem{Berlin:2016vnh}
A.~Berlin, D.~Hooper and G.~Krnjaic, ``{PeV-Scale Dark Matter as a Thermal
  Relic of a Decoupled
  Sector},''\href{http://dx.doi.org/10.1016/j.physletb.2016.06.037}{\emph{Phys.
  Lett. B} {\bf 760} (2016) 106--111},
  [\href{https://arxiv.org/abs/1602.08490}{{\tt 1602.08490}}].

\bibitem{Berlin:2016gtr}
A.~Berlin, D.~Hooper and G.~Krnjaic, ``{Thermal Dark Matter From A Highly
  Decoupled
  Sector},''\href{http://dx.doi.org/10.1103/PhysRevD.94.095019}{\emph{Phys.
  Rev. D} {\bf 94} (2016) 095019},
  [\href{https://arxiv.org/abs/1609.02555}{{\tt 1609.02555}}].

\bibitem{Bramante:2017obj}
J.~Bramante and J.~Unwin, ``{Superheavy Thermal Dark Matter and Primordial
  Asymmetries},''\href{http://dx.doi.org/10.1007/JHEP02(2017)119}{\emph{JHEP}
  {\bf 02} (2017) 119}, [\href{https://arxiv.org/abs/1701.05859}{{\tt
  1701.05859}}].

\bibitem{Hamdan:2017psw}
S.~Hamdan and J.~Unwin, ``{Dark Matter Freeze-out During Matter
  Domination},''\href{http://dx.doi.org/10.1142/S021773231850181X}{\emph{Mod.
  Phys. Lett. A} {\bf 33} (2018) 1850181},
  [\href{https://arxiv.org/abs/1710.03758}{{\tt 1710.03758}}].

\bibitem{Cirelli:2018iax}
M.~Cirelli, Y.~Gouttenoire, K.~Petraki and F.~Sala, ``{Homeopathic Dark Matter,
  or how diluted heavy substances produce high energy cosmic
  rays},''\href{http://dx.doi.org/10.1088/1475-7516/2019/02/014}{\emph{JCAP}
  {\bf 02} (2019) 014}, [\href{https://arxiv.org/abs/1811.03608}{{\tt
  1811.03608}}].

\bibitem{Babichev:2018mtd}
E.~Babichev, D.~Gorbunov and S.~Ramazanov, ``{New mechanism of producing
  superheavy Dark
  Matter},''\href{http://dx.doi.org/10.1016/j.physletb.2019.05.030}{\emph{Phys.
  Lett. B} {\bf 794} (2019) 69--76},
  [\href{https://arxiv.org/abs/1812.03516}{{\tt 1812.03516}}].

\bibitem{Hashiba:2018tbu}
S.~Hashiba and J.~Yokoyama, ``{Gravitational particle creation for dark matter
  and
  reheating},''\href{http://dx.doi.org/10.1103/PhysRevD.99.043008}{\emph{Phys.
  Rev. D} {\bf 99} (2019) 043008},
  [\href{https://arxiv.org/abs/1812.10032}{{\tt 1812.10032}}].

\bibitem{Hooper:2019gtx}
D.~Hooper, G.~Krnjaic and S.~D. McDermott, ``{Dark Radiation and Superheavy
  Dark Matter from Black Hole
  Domination},''\href{http://dx.doi.org/10.1007/JHEP08(2019)001}{\emph{JHEP}
  {\bf 08} (2019) 001}, [\href{https://arxiv.org/abs/1905.01301}{{\tt
  1905.01301}}].

\bibitem{Davoudiasl:2019xeb}
H.~Davoudiasl and G.~Mohlabeng, ``{Getting a THUMP from a
  WIMP},''\href{http://dx.doi.org/10.1007/JHEP04(2020)177}{\emph{JHEP} {\bf 04}
  (2020) 177}, [\href{https://arxiv.org/abs/1912.05572}{{\tt 1912.05572}}].

\bibitem{Chanda:2019xyl}
P.~Chanda, S.~Hamdan and J.~Unwin, ``{Reviving $Z$ and Higgs Mediated Dark
  Matter Models in Matter Dominated
  Freeze-out},''\href{http://dx.doi.org/10.1088/1475-7516/2020/01/034}{\emph{JCAP}
  {\bf 01} (2020) 034}, [\href{https://arxiv.org/abs/1911.02616}{{\tt
  1911.02616}}].

\bibitem{Baker:2019ndr}
M.~J. Baker, J.~Kopp and A.~J. Long, ``{Filtered Dark Matter at a First Order
  Phase
  Transition},''\href{http://dx.doi.org/10.1103/PhysRevLett.125.151102}{\emph{Phys.
  Rev. Lett.} {\bf 125} (2020) 151102},
  [\href{https://arxiv.org/abs/1912.02830}{{\tt 1912.02830}}].

\bibitem{Chway:2019kft}
D.~Chway, T.~H. Jung and C.~S. Shin, ``{Dark matter filtering-out effect during
  a first-order phase
  transition},''\href{http://dx.doi.org/10.1103/PhysRevD.101.095019}{\emph{Phys.
  Rev. D} {\bf 101} (2020) 095019},
  [\href{https://arxiv.org/abs/1912.04238}{{\tt 1912.04238}}].

\bibitem{Marfatia:2020bcs}
D.~Marfatia and P.-Y. Tseng, ``{Gravitational wave signals of dark matter
  freeze-out},''\href{http://dx.doi.org/10.1007/JHEP02(2021)022}{\emph{JHEP}
  {\bf 02} (2021) 022}, [\href{https://arxiv.org/abs/2006.07313}{{\tt
  2006.07313}}].

\bibitem{Baldes:2020kam}
I.~Baldes, Y.~Gouttenoire and F.~Sala, ``{String Fragmentation in Supercooled
  Confinement and Implications for Dark
  Matter},''\href{http://dx.doi.org/10.1007/JHEP04(2021)278}{\emph{JHEP} {\bf
  04} (2021) 278}, [\href{https://arxiv.org/abs/2007.08440}{{\tt 2007.08440}}].

\bibitem{Azatov:2021ifm}
A.~Azatov, M.~Vanvlasselaer and W.~Yin, ``{Dark Matter production from
  relativistic bubble
  walls},''\href{http://dx.doi.org/10.1007/JHEP03(2021)288}{\emph{JHEP} {\bf
  03} (2021) 288}, [\href{https://arxiv.org/abs/2101.05721}{{\tt 2101.05721}}].

\bibitem{Berlin:2017ife}
A.~Berlin, ``{WIMPs with GUTs: Dark Matter Coannihilation with a Lighter
  Species},''\href{http://dx.doi.org/10.1103/PhysRevLett.119.121801}{\emph{Phys.
  Rev. Lett.} {\bf 119} (2017) 121801},
  [\href{https://arxiv.org/abs/1704.08256}{{\tt 1704.08256}}].

\bibitem{Kramer:2020sbb}
E.~D. Kramer, E.~Kuflik, N.~Levi, N.~J. Outmezguine and J.~T. Ruderman,
  ``{Heavy Thermal Dark Matter from a New Collision
  Mechanism},''\href{http://dx.doi.org/10.1103/PhysRevLett.126.081802}{\emph{Phys.
  Rev. Lett.} {\bf 126} (2021) 081802},
  [\href{https://arxiv.org/abs/2003.04900}{{\tt 2003.04900}}].

\bibitem{Davidson:1978pm}
A.~Davidson, ``{$B-L$ as the fourth color within an $\mathrm{SU}(2)_L \times
  \mathrm{U}(1)_R \times \mathrm{U}(1)$
  model},''\href{http://dx.doi.org/10.1103/PhysRevD.20.776}{\emph{Phys. Rev. D}
  {\bf 20} (1979) 776}.

\bibitem{Marshak:1979fm}
R.~E. Marshak and R.~N. Mohapatra, ``{Quark - Lepton Symmetry and B-L as the
  U(1) Generator of the Electroweak Symmetry
  Group},''\href{http://dx.doi.org/10.1016/0370-2693(80)90436-0}{\emph{Phys.
  Lett. B} {\bf 91} (1980) 222--224}.

\bibitem{Mohapatra:1980qe}
R.~N. Mohapatra and R.~E. Marshak, ``{Local B-L Symmetry of Electroweak
  Interactions, Majorana Neutrinos and Neutron
  Oscillations},''\href{http://dx.doi.org/10.1103/PhysRevLett.44.1316}{\emph{Phys.
  Rev. Lett.} {\bf 44} (1980) 1316--1319}.

\bibitem{Davidson:1987mh}
A.~Davidson and K.~C. Wali, ``{Universal Seesaw
  Mechanism?},''\href{http://dx.doi.org/10.1103/PhysRevLett.59.393}{\emph{Phys.
  Rev. Lett.} {\bf 59} (1987) 393}.

\bibitem{Minkowski:1977sc}
P.~Minkowski, ``{$\mu \to e\gamma$ at a Rate of One Out of $10^{9}$ Muon
  Decays?},''\href{http://dx.doi.org/10.1016/0370-2693(77)90435-X}{\emph{Phys.
  Lett. B} {\bf 67} (1977) 421--428}.

\bibitem{Fukugita:1986hr}
M.~Fukugita and T.~Yanagida, ``{Baryogenesis Without Grand
  Unification},''\href{http://dx.doi.org/10.1016/0370-2693(86)91126-3}{\emph{Phys.
  Lett. B} {\bf 174} (1986) 45--47}.

\bibitem{KATRIN:2019yun}
{\scshape KATRIN} collaboration, M.~Aker et~al., ``{Improved Upper Limit on the
  Neutrino Mass from a Direct Kinematic Method by
  KATRIN},''\href{http://dx.doi.org/10.1103/PhysRevLett.123.221802}{\emph{Phys.
  Rev. Lett.} {\bf 123} (2019) 221802},
  [\href{https://arxiv.org/abs/1909.06048}{{\tt 1909.06048}}].

\bibitem{Okada:2010wd}
N.~Okada and O.~Seto, ``{Higgs portal dark matter in the minimal gauged
  $U(1)_{B-L}$
  model},''\href{http://dx.doi.org/10.1103/PhysRevD.82.023507}{\emph{Phys. Rev.
  D} {\bf 82} (2010) 023507}, [\href{https://arxiv.org/abs/1002.2525}{{\tt
  1002.2525}}].

\bibitem{Okada:2012sg}
N.~Okada and Y.~Orikasa, ``{Dark matter in the classically conformal B-L
  model},''\href{http://dx.doi.org/10.1103/PhysRevD.85.115006}{\emph{Phys. Rev.
  D} {\bf 85} (2012) 115006}, [\href{https://arxiv.org/abs/1202.1405}{{\tt
  1202.1405}}].

\bibitem{Okada:2016tci}
N.~Okada and S.~Okada, ``{$Z^\prime$-portal right-handed neutrino dark matter
  in the minimal U(1)$_X$ extended Standard
  Model},''\href{http://dx.doi.org/10.1103/PhysRevD.95.035025}{\emph{Phys. Rev.
  D} {\bf 95} (2017) 035025}, [\href{https://arxiv.org/abs/1611.02672}{{\tt
  1611.02672}}].

\bibitem{Okada:2016gsh}
N.~Okada and S.~Okada, ``{$Z^\prime_{BL}$ portal dark matter and LHC Run-2
  results},''\href{http://dx.doi.org/10.1103/PhysRevD.93.075003}{\emph{Phys.
  Rev. D} {\bf 93} (2016) 075003},
  [\href{https://arxiv.org/abs/1601.07526}{{\tt 1601.07526}}].

\bibitem{Okada:2018ktp}
S.~Okada, ``{$Z'$ Portal Dark Matter in the Minimal $B-L$
  Model},''\href{http://dx.doi.org/10.1155/2018/5340935}{\emph{Adv. High Energy
  Phys.} {\bf 2018} (2018) 5340935},
  [\href{https://arxiv.org/abs/1803.06793}{{\tt 1803.06793}}].

\bibitem{Borah:2018smz}
D.~Borah, D.~Nanda, N.~Narendra and N.~Sahu, ``{Right-handed neutrino dark
  matter with radiative neutrino mass in gauged B \ensuremath{-} L
  model},''\href{http://dx.doi.org/10.1016/j.nuclphysb.2019.114841}{\emph{Nucl.
  Phys. B} {\bf 950} (2020) 114841},
  [\href{https://arxiv.org/abs/1810.12920}{{\tt 1810.12920}}].

\bibitem{Xing:2020ald}
Z.-z. Xing and Z.-h. Zhao, ``{The minimal seesaw and leptogenesis
  models},''\href{http://dx.doi.org/10.1088/1361-6633/abf086}{\emph{Rept. Prog.
  Phys.} {\bf 84} (2021) 066201}, [\href{https://arxiv.org/abs/2008.12090}{{\tt
  2008.12090}}].

\bibitem{Bian:2019szo}
L.~Bian, W.~Cheng, H.-K. Guo and Y.~Zhang, ``{Gravitational waves triggered by
  $B-L$ charged hidden scalar and leptogenesis},''
  \href{https://arxiv.org/abs/1907.13589}{{\tt 1907.13589}}.

\bibitem{Cirelli:2012ut}
M.~Cirelli, E.~Moulin, P.~Panci, P.~D. Serpico and A.~Viana, ``{Gamma ray
  constraints on Decaying Dark
  Matter},''\href{http://dx.doi.org/10.1103/PhysRevD.86.083506}{\emph{Phys.
  Rev. D} {\bf 86} (2012) 083506}, [\href{https://arxiv.org/abs/1205.5283}{{\tt
  1205.5283}}].

\bibitem{Essig:2013goa}
R.~Essig, E.~Kuflik, S.~D. McDermott, T.~Volansky and K.~M. Zurek,
  ``{Constraining Light Dark Matter with Diffuse X-Ray and Gamma-Ray
  Observations},''\href{http://dx.doi.org/10.1007/JHEP11(2013)193}{\emph{JHEP}
  {\bf 11} (2013) 193}, [\href{https://arxiv.org/abs/1309.4091}{{\tt
  1309.4091}}].

\bibitem{Blanco:2018esa}
C.~Blanco and D.~Hooper, ``{Constraints on Decaying Dark Matter from the
  Isotropic Gamma-Ray
  Background},''\href{http://dx.doi.org/10.1088/1475-7516/2019/03/019}{\emph{JCAP}
  {\bf 03} (2019) 019}, [\href{https://arxiv.org/abs/1811.05988}{{\tt
  1811.05988}}].

\bibitem{kolb1981early}
E.~W. Kolb and M.~S. Turner, ``The early universe,''{\emph{Nature} {\bf 294}
  (1981) 521--526}.

\bibitem{Alloul:2013bka}
A.~Alloul, N.~D. Christensen, C.~Degrande, C.~Duhr and B.~Fuks, ``{FeynRules
  2.0 - A complete toolbox for tree-level
  phenomenology},''\href{http://dx.doi.org/10.1016/j.cpc.2014.04.012}{\emph{Comput.
  Phys. Commun.} {\bf 185} (2014) 2250--2300},
  [\href{https://arxiv.org/abs/1310.1921}{{\tt 1310.1921}}].

\bibitem{Alwall:2014hca}
J.~Alwall, R.~Frederix, S.~Frixione, V.~Hirschi, F.~Maltoni, O.~Mattelaer
  et~al., ``{The automated computation of tree-level and next-to-leading order
  differential cross sections, and their matching to parton shower
  simulations},''\href{http://dx.doi.org/10.1007/JHEP07(2014)079}{\emph{JHEP}
  {\bf 07} (2014) 079}, [\href{https://arxiv.org/abs/1405.0301}{{\tt
  1405.0301}}].

\bibitem{Buchmuller:2013lra}
W.~Buchm\"uller, V.~Domcke, K.~Kamada and K.~Schmitz, ``{The Gravitational Wave
  Spectrum from Cosmological $B-L$
  Breaking},''\href{http://dx.doi.org/10.1088/1475-7516/2013/10/003}{\emph{JCAP}
  {\bf 10} (2013) 003}, [\href{https://arxiv.org/abs/1305.3392}{{\tt
  1305.3392}}].

\bibitem{Dror:2019syi}
J.~A. Dror, T.~Hiramatsu, K.~Kohri, H.~Murayama and G.~White, ``{Testing the
  Seesaw Mechanism and Leptogenesis with Gravitational
  Waves},''\href{http://dx.doi.org/10.1103/PhysRevLett.124.041804}{\emph{Phys.
  Rev. Lett.} {\bf 124} (2020) 041804},
  [\href{https://arxiv.org/abs/1908.03227}{{\tt 1908.03227}}].

\bibitem{Auclair:2019wcv}
P.~Auclair et~al., ``{Probing the gravitational wave background from cosmic
  strings with
  LISA},''\href{http://dx.doi.org/10.1088/1475-7516/2020/04/034}{\emph{JCAP}
  {\bf 04} (2020) 034}, [\href{https://arxiv.org/abs/1909.00819}{{\tt
  1909.00819}}].

\bibitem{Fornal:2020esl}
B.~Fornal and B.~Shams Es~Haghi, ``{Baryon and Lepton Number Violation from
  Gravitational
  Waves},''\href{http://dx.doi.org/10.1103/PhysRevD.102.115037}{\emph{Phys.
  Rev. D} {\bf 102} (2020) 115037},
  [\href{https://arxiv.org/abs/2008.05111}{{\tt 2008.05111}}].

\bibitem{Samanta:2020cdk}
R.~Samanta and S.~Datta, ``{Gravitational wave complementarity and impact of
  NANOGrav data on gravitational
  leptogenesis},''\href{http://dx.doi.org/10.1007/JHEP05(2021)211}{\emph{JHEP}
  {\bf 05} (2021) 211}, [\href{https://arxiv.org/abs/2009.13452}{{\tt
  2009.13452}}].

\bibitem{Masoud:2021prr}
M.~A. Masoud, M.~U. Rehman and Q.~Shafi, ``{Sneutrino Tribrid Inflation,
  Metastable Cosmic Strings and Gravitational Waves},''
  \href{https://arxiv.org/abs/2107.09689}{{\tt 2107.09689}}.

\bibitem{Buchmuller:2021mbb}
W.~Buchmuller, V.~Domcke and K.~Schmitz, ``{Stochastic gravitational-wave
  background from metastable cosmic strings},''
  \href{https://arxiv.org/abs/2107.04578}{{\tt 2107.04578}}.

\bibitem{Nielsen:1973cs}
H.~B. Nielsen and P.~Olesen, ``{Vortex Line Models for Dual
  Strings},''\href{http://dx.doi.org/10.1016/0550-3213(73)90350-7}{\emph{Nucl.
  Phys. B} {\bf 61} (1973) 45--61}.

\bibitem{Vachaspati:1984gt}
T.~Vachaspati and A.~Vilenkin, ``{Gravitational Radiation from Cosmic
  Strings},''\href{http://dx.doi.org/10.1103/PhysRevD.31.3052}{\emph{Phys. Rev.
  D} {\bf 31} (1985) 3052}.

\bibitem{Blanco-Pillado:2017oxo}
J.~J. Blanco-Pillado and K.~D. Olum, ``{Stochastic gravitational wave
  background from smoothed cosmic string
  loops},''\href{http://dx.doi.org/10.1103/PhysRevD.96.104046}{\emph{Phys. Rev.
  D} {\bf 96} (2017) 104046}, [\href{https://arxiv.org/abs/1709.02693}{{\tt
  1709.02693}}].

\bibitem{Binetruy:2012ze}
P.~Binetruy, A.~Bohe, C.~Caprini and J.-F. Dufaux, ``{Cosmological Backgrounds
  of Gravitational Waves and eLISA/NGO: Phase Transitions, Cosmic Strings and
  Other
  Sources},''\href{http://dx.doi.org/10.1088/1475-7516/2012/06/027}{\emph{JCAP}
  {\bf 06} (2012) 027}, [\href{https://arxiv.org/abs/1201.0983}{{\tt
  1201.0983}}].

\bibitem{Blanco-Pillado:2013qja}
J.~J. Blanco-Pillado, K.~D. Olum and B.~Shlaer, ``{The number of cosmic string
  loops},''\href{http://dx.doi.org/10.1103/PhysRevD.89.023512}{\emph{Phys. Rev.
  D} {\bf 89} (2014) 023512}, [\href{https://arxiv.org/abs/1309.6637}{{\tt
  1309.6637}}].

\bibitem{McLaughlin:2013ira}
M.~A. McLaughlin, ``{The North American Nanohertz Observatory for Gravitational
  Waves},''\href{http://dx.doi.org/10.1088/0264-9381/30/22/224008}{\emph{Class.
  Quant. Grav.} {\bf 30} (2013) 224008},
  [\href{https://arxiv.org/abs/1310.0758}{{\tt 1310.0758}}].

\bibitem{NANOGRAV:2018hou}
{\scshape NANOGRAV} collaboration, Z.~Arzoumanian et~al., ``{The NANOGrav
  11-year Data Set: Pulsar-timing Constraints On The Stochastic
  Gravitational-wave
  Background},''\href{http://dx.doi.org/10.3847/1538-4357/aabd3b}{\emph{Astrophys.
  J.} {\bf 859} (2018) 47}, [\href{https://arxiv.org/abs/1801.02617}{{\tt
  1801.02617}}].

\bibitem{Aggarwal:2018mgp}
K.~Aggarwal et~al., ``{The NANOGrav 11-Year Data Set: Limits on Gravitational
  Waves from Individual Supermassive Black Hole
  Binaries},''\href{http://dx.doi.org/10.3847/1538-4357/ab2236}{\emph{Astrophys.
  J.} {\bf 880} (2019) 2}, [\href{https://arxiv.org/abs/1812.11585}{{\tt
  1812.11585}}].

\bibitem{Brazier:2019mmu}
A.~Brazier et~al., ``{The NANOGrav Program for Gravitational Waves and
  Fundamental Physics},'' \href{https://arxiv.org/abs/1908.05356}{{\tt
  1908.05356}}.

\bibitem{Manchester:2012za}
R.~N. Manchester et~al., ``{The Parkes Pulsar Timing Array
  Project},''\href{http://dx.doi.org/10.1017/pasa.2012.017}{\emph{Publ. Astron.
  Soc. Austral.} {\bf 30} (2013) 17},
  [\href{https://arxiv.org/abs/1210.6130}{{\tt 1210.6130}}].

\bibitem{Shannon:2015ect}
R.~M. Shannon et~al., ``{Gravitational waves from binary supermassive black
  holes missing in pulsar
  observations},''\href{http://dx.doi.org/10.1126/science.aab1910}{\emph{Science}
  {\bf 349} (2015) 1522--1525}, [\href{https://arxiv.org/abs/1509.07320}{{\tt
  1509.07320}}].

\bibitem{Kramer:2013kea}
M.~Kramer and D.~J. Champion, ``{The European Pulsar Timing Array and the Large
  European Array for
  Pulsars},''\href{http://dx.doi.org/10.1088/0264-9381/30/22/224009}{\emph{Class.
  Quant. Grav.} {\bf 30} (2013) 224009}.

\bibitem{Lentati:2015qwp}
L.~Lentati et~al., ``{European Pulsar Timing Array Limits On An Isotropic
  Stochastic Gravitational-Wave
  Background},''\href{http://dx.doi.org/10.1093/mnras/stv1538}{\emph{Mon. Not.
  Roy. Astron. Soc.} {\bf 453} (2015) 2576--2598},
  [\href{https://arxiv.org/abs/1504.03692}{{\tt 1504.03692}}].

\bibitem{Babak:2015lua}
S.~Babak et~al., ``{European Pulsar Timing Array Limits on Continuous
  Gravitational Waves from Individual Supermassive Black Hole
  Binaries},''\href{http://dx.doi.org/10.1093/mnras/stv2092}{\emph{Mon. Not.
  Roy. Astron. Soc.} {\bf 455} (2016) 1665--1679},
  [\href{https://arxiv.org/abs/1509.02165}{{\tt 1509.02165}}].

\bibitem{Hobbs:2009yy}
G.~Hobbs et~al., ``{The international pulsar timing array project: using
  pulsars as a gravitational wave
  detector},''\href{http://dx.doi.org/10.1088/0264-9381/27/8/084013}{\emph{Class.
  Quant. Grav.} {\bf 27} (2010) 084013},
  [\href{https://arxiv.org/abs/0911.5206}{{\tt 0911.5206}}].

\bibitem{Manchester:2013ndt}
R.~N. Manchester, ``{The International Pulsar Timing
  Array},''\href{http://dx.doi.org/10.1088/0264-9381/30/22/224010}{\emph{Class.
  Quant. Grav.} {\bf 30} (2013) 224010},
  [\href{https://arxiv.org/abs/1309.7392}{{\tt 1309.7392}}].

\bibitem{Verbiest:2016vem}
J.~P.~W. Verbiest et~al., ``{The International Pulsar Timing Array: First Data
  Release},''\href{http://dx.doi.org/10.1093/mnras/stw347}{\emph{Mon. Not. Roy.
  Astron. Soc.} {\bf 458} (2016) 1267--1288},
  [\href{https://arxiv.org/abs/1602.03640}{{\tt 1602.03640}}].

\bibitem{Hazboun:2018wpv}
J.~S. Hazboun, C.~M.~F. Mingarelli and K.~Lee, ``{The Second International
  Pulsar Timing Array Mock Data Challenge},''
  \href{https://arxiv.org/abs/1810.10527}{{\tt 1810.10527}}.

\bibitem{Carilli:2004nx}
C.~L. Carilli and S.~Rawlings, ``{Science with the Square Kilometer Array:
  Motivation, key science projects, standards and
  assumptions},''\href{http://dx.doi.org/10.1016/j.newar.2004.09.001}{\emph{New
  Astron. Rev.} {\bf 48} (2004) 979},
  [\href{https://arxiv.org/abs/astro-ph/0409274}{{\tt astro-ph/0409274}}].

\bibitem{Janssen:2014dka}
G.~Janssen et~al., ``{Gravitational wave astronomy with the
  SKA},''\href{http://dx.doi.org/10.22323/1.215.0037}{\emph{PoS} {\bf AASKA14}
  (2015) 037}, [\href{https://arxiv.org/abs/1501.00127}{{\tt 1501.00127}}].

\bibitem{Weltman:2018zrl}
A.~Weltman et~al., ``{Fundamental physics with the Square Kilometre
  Array},''\href{http://dx.doi.org/10.1017/pasa.2019.42}{\emph{Publ. Astron.
  Soc. Austral.} {\bf 37} (2020) e002},
  [\href{https://arxiv.org/abs/1810.02680}{{\tt 1810.02680}}].

\bibitem{LISA:2017pwj}
{\scshape LISA} collaboration, P.~Amaro-Seoane et~al., ``{Laser Interferometer
  Space Antenna},'' \href{https://arxiv.org/abs/1702.00786}{{\tt 1702.00786}}.

\bibitem{Crowder:2005nr}
J.~Crowder and N.~J. Cornish, ``{Beyond LISA: Exploring future gravitational
  wave
  missions},''\href{http://dx.doi.org/10.1103/PhysRevD.72.083005}{\emph{Phys.
  Rev. D} {\bf 72} (2005) 083005},
  [\href{https://arxiv.org/abs/gr-qc/0506015}{{\tt gr-qc/0506015}}].

\bibitem{TianQin:2015yph}
{\scshape TianQin} collaboration, J.~Luo et~al., ``{TianQin: a space-borne
  gravitational wave
  detector},''\href{http://dx.doi.org/10.1088/0264-9381/33/3/035010}{\emph{Class.
  Quant. Grav.} {\bf 33} (2016) 035010},
  [\href{https://arxiv.org/abs/1512.02076}{{\tt 1512.02076}}].

\bibitem{Hu:2017yoc}
Y.-M. Hu, J.~Mei and J.~Luo, ``{Science prospects for space-borne
  gravitational-wave
  missions},''\href{http://dx.doi.org/10.1093/nsr/nwx115}{\emph{Natl. Sci.
  Rev.} {\bf 4} (2017) 683--684}.

\bibitem{TianQin:2020hid}
{\scshape TianQin} collaboration, J.~Mei et~al., ``{The TianQin project:
  current progress on science and technology},''
  \href{https://arxiv.org/abs/2008.10332}{{\tt 2008.10332}}.

\bibitem{Hu:2017mde}
W.-R. Hu and Y.-L. Wu, ``{The Taiji Program in Space for gravitational wave
  physics and the nature of
  gravity},''\href{http://dx.doi.org/10.1093/nsr/nwx116}{\emph{Natl. Sci. Rev.}
  {\bf 4} (2017) 685--686}.

\bibitem{Ruan:2018tsw}
W.-H. Ruan, Z.-K. Guo, R.-G. Cai and Y.-Z. Zhang, ``{Taiji program:
  Gravitational-wave
  sources},''\href{http://dx.doi.org/10.1142/S0217751X2050075X}{\emph{Int. J.
  Mod. Phys. A} {\bf 35} (2020) 2050075},
  [\href{https://arxiv.org/abs/1807.09495}{{\tt 1807.09495}}].

\bibitem{LIGOScientific:2014qfs}
{\scshape LIGO Scientific, VIRGO} collaboration, J.~Aasi et~al.,
  ``{Characterization of the LIGO detectors during their sixth science
  run},''\href{http://dx.doi.org/10.1088/0264-9381/32/11/115012}{\emph{Class.
  Quant. Grav.} {\bf 32} (2015) 115012},
  [\href{https://arxiv.org/abs/1410.7764}{{\tt 1410.7764}}].

\bibitem{LIGOScientific:2019vic}
{\scshape LIGO Scientific, Virgo} collaboration, B.~P. Abbott et~al., ``{Search
  for the isotropic stochastic background using data from Advanced
  LIGO\textquoteright{}s second observing
  run},''\href{http://dx.doi.org/10.1103/PhysRevD.100.061101}{\emph{Phys. Rev.
  D} {\bf 100} (2019) 061101}, [\href{https://arxiv.org/abs/1903.02886}{{\tt
  1903.02886}}].

\bibitem{Reitze:2019iox}
D.~Reitze et~al., ``{Cosmic Explorer: The U.S. Contribution to
  Gravitational-Wave Astronomy beyond LIGO},''{\emph{Bull. Am. Astron. Soc.}
  {\bf 51} (2019) 035}, [\href{https://arxiv.org/abs/1907.04833}{{\tt
  1907.04833}}].

\bibitem{Punturo:2010zz}
M.~Punturo et~al., ``{The Einstein Telescope: A third-generation gravitational
  wave
  observatory},''\href{http://dx.doi.org/10.1088/0264-9381/27/19/194002}{\emph{Class.
  Quant. Grav.} {\bf 27} (2010) 194002}.

\bibitem{Hild:2010id}
S.~Hild et~al., ``{Sensitivity Studies for Third-Generation Gravitational Wave
  Observatories},''\href{http://dx.doi.org/10.1088/0264-9381/28/9/094013}{\emph{Class.
  Quant. Grav.} {\bf 28} (2011) 094013},
  [\href{https://arxiv.org/abs/1012.0908}{{\tt 1012.0908}}].

\bibitem{Sathyaprakash:2012jk}
B.~Sathyaprakash et~al., ``{Scientific Objectives of Einstein
  Telescope},''\href{http://dx.doi.org/10.1088/0264-9381/29/12/124013}{\emph{Class.
  Quant. Grav.} {\bf 29} (2012) 124013},
  [\href{https://arxiv.org/abs/1206.0331}{{\tt 1206.0331}}].

\bibitem{Schmitz:2020syl}
K.~Schmitz, ``{New Sensitivity Curves for Gravitational-Wave Signals from
  Cosmological Phase
  Transitions},''\href{http://dx.doi.org/10.1007/JHEP01(2021)097}{\emph{JHEP}
  {\bf 01} (2021) 097}, [\href{https://arxiv.org/abs/2002.04615}{{\tt
  2002.04615}}].

\bibitem{Ringeval:2017eww}
C.~Ringeval and T.~Suyama, ``{Stochastic gravitational waves from cosmic string
  loops in
  scaling},''\href{http://dx.doi.org/10.1088/1475-7516/2017/12/027}{\emph{JCAP}
  {\bf 12} (2017) 027}, [\href{https://arxiv.org/abs/1709.03845}{{\tt
  1709.03845}}].

\bibitem{Blanco-Pillado:2017rnf}
J.~J. Blanco-Pillado, K.~D. Olum and X.~Siemens, ``{New limits on cosmic
  strings from gravitational wave
  observation},''\href{http://dx.doi.org/10.1016/j.physletb.2018.01.050}{\emph{Phys.
  Lett. B} {\bf 778} (2018) 392--396},
  [\href{https://arxiv.org/abs/1709.02434}{{\tt 1709.02434}}].

\bibitem{NANOGrav:2020bcs}
{\scshape NANOGrav} collaboration, Z.~Arzoumanian et~al., ``{The NANOGrav 12.5
  yr Data Set: Search for an Isotropic Stochastic Gravitational-wave
  Background},''\href{http://dx.doi.org/10.3847/2041-8213/abd401}{\emph{Astrophys.
  J. Lett.} {\bf 905} (2020) L34},
  [\href{https://arxiv.org/abs/2009.04496}{{\tt 2009.04496}}].

\bibitem{Ellis:2020ena}
J.~Ellis and M.~Lewicki, ``{Cosmic String Interpretation of NANOGrav Pulsar
  Timing
  Data},''\href{http://dx.doi.org/10.1103/PhysRevLett.126.041304}{\emph{Phys.
  Rev. Lett.} {\bf 126} (2021) 041304},
  [\href{https://arxiv.org/abs/2009.06555}{{\tt 2009.06555}}].

\bibitem{Bian:2020urb}
L.~Bian, R.-G. Cai, J.~Liu, X.-Y. Yang and R.~Zhou, ``{Evidence for different
  gravitational-wave sources in the NANOGrav
  dataset},''\href{http://dx.doi.org/10.1103/PhysRevD.103.L081301}{\emph{Phys.
  Rev. D} {\bf 103} (2021) L081301},
  [\href{https://arxiv.org/abs/2009.13893}{{\tt 2009.13893}}].

\bibitem{Blasi:2020mfx}
S.~Blasi, V.~Brdar and K.~Schmitz, ``{Has NANOGrav found first evidence for
  cosmic
  strings?},''\href{http://dx.doi.org/10.1103/PhysRevLett.126.041305}{\emph{Phys.
  Rev. Lett.} {\bf 126} (2021) 041305},
  [\href{https://arxiv.org/abs/2009.06607}{{\tt 2009.06607}}].

\bibitem{LIGOScientific:2021nrg}
{\scshape LIGO Scientific, Virgo, KAGRA} collaboration, R.~Abbott et~al.,
  ``{Constraints on Cosmic Strings Using Data from the Third Advanced
  LIGO\textendash{}Virgo Observing
  Run},''\href{http://dx.doi.org/10.1103/PhysRevLett.126.241102}{\emph{Phys.
  Rev. Lett.} {\bf 126} (2021) 241102},
  [\href{https://arxiv.org/abs/2101.12248}{{\tt 2101.12248}}].

\bibitem{Giudice:2003jh}
G.~F. Giudice, A.~Notari, M.~Raidal, A.~Riotto and A.~Strumia, ``{Towards a
  complete theory of thermal leptogenesis in the SM and
  MSSM},''\href{http://dx.doi.org/10.1016/j.nuclphysb.2004.02.019}{\emph{Nucl.
  Phys. B} {\bf 685} (2004) 89--149},
  [\href{https://arxiv.org/abs/hep-ph/0310123}{{\tt hep-ph/0310123}}].

\end{thebibliography}\endgroup

\end{document}